\documentclass[twocolumn,prb]{revtex4}
\usepackage{graphicx}    
\usepackage{dcolumn}     
\usepackage{slashbox}     
\usepackage{bm}          
\newcommand{\Vec}[1]{\mbox{\boldmath$#1$}}
\begin{document}
\title{Pnictogen height as a possible switch between 
high-$T_c$ nodeless and low-$T_c$ nodal pairings  
in the iron based superconductors}

\author{Kazuhiko Kuroki$^{1,5}$, Hidetomo Usui$^{1}$, 
Seiichiro Onari$^{2,5}$, Ryotaro Arita$^{3,5,6}$, and 
Hideo Aoki$^{4,5}$ }

\affiliation{
$^1$ Department of Applied Physics and Chemistry, 
The University of Electro -Communications, Chofu, Tokyo 182-8585, Japan\\
$^2$ Department of Applied Physics, Nagoya University, 
Nagoya 464-8603, Japan\\
$^3$ Department of Applied Physics,  
University of Tokyo, Hongo, Tokyo 113-8656, Japan\\
$^4$ Department of Physics,  
University of Tokyo, Hongo, Tokyo 113-0033, Japan\\
$^5$ JST, TRIP, Sanbancho, Chiyoda, Tokyo 102-0075, Japan\\
$^6$ JST, CREST, Hongo, Tokyo 113-8656, Japan}

\begin{abstract}
We study the effect of the lattice structure on the 
spin-fluctuation mediated superconductivity in the iron pnictides 
adopting the five-band models of several virtual lattice structures of LaFeAsO 
as well as actual materials such as NdFeAsO and LaFePO 
obtained from the maximally-localized Wannier orbitals.  
Random phase approximation is applied to the models to solve the 
Eliashberg equation.  This reveals that 
the gap function and the strength of the superconducting instability 
are determined by the cooperation or competition among 
multiple spin fluctuation modes arising from several nestings among 
disconnected pieces of the Fermi surface, which is affected by the 
lattice structure. Specifically, the appearance of the 
Fermi surface $\gamma$ around $(\pi,\pi)$ in the unfolded Brillouin zone 
is  sensitive to the pnictogen height $h_{\rm Pn}$ measured from the Fe plane, 
where $h_{\rm Pn}$ is shown to act as a switch between high-$T_c$ nodeless 
and low-$T_c$ nodal pairings.  We also find that reduction in the 
lattice constants generally suppresses superconductivity.   
We can then combine these to obtain a generic superconducting phase diagram 
against the pnictogen height and lattice constant.  This 
suggests that 
NdFeAsO is expected to exhibit a fully-gapped, sign-reversing $s$-wave 
superconductivity with a higher $T_c$ than in LaFeAsO, while a nodal pairing 
with a low $T_c$ is expected for LaFePO, which is consistent with experiments. 
\end{abstract}
\pacs{PACS numbers: }
\maketitle

\section{Introduction}
The discovery of superconductivity in the 
iron-based compounds by Hosono's group\cite{Hosono} 
and subsequent increase in the 
transition temperature ($T_c$) exceeding 50 K\cite{Ren} 
in the same family of compounds 
are seminal not only because of high values of $T_c$, 
but also because this poses a fundamental question 
on electronic mechanisms of high $T_c$ superconductivity 
in a wider class of compounds other than cuprates.

Theoretically, a phonon mechanism was 
shown to be unlikely for this system,\cite{Boeri} and a 
spin-fluctuation mediated pairing has been 
proposed from the very early stage of the study.\cite{Mazin,Kuroki1st,Mazinrev}
In these studies the nesting between disconnected pieces 
(pockets) of the Fermi surface is shown to 
induce spin fluctuations associated with the nesting vector.  This 
can give rise to a superconducting gap, which is basically 
$s$-wave but changes sign between different pockets, 
hence termed as $s\pm$wave 
or sign-reversing $s$-wave first proposed by Mazin {\it et al.}
\cite{Mazin}(see Fig.\ref{fig4}).  
Although recent experimental as well as theoretical 
studies suggest that the magnetism in the {\it undoped} material is not 
driven solely by Fermi surface nesting,\cite{Mazinrev} 
the spin fluctuation originating from the nesting has been considered 
to be a possible origin of the pairing interaction by a number of 
authors.\cite{Kuroki1st,errata,Qi,Raghu,Daghofer,Chubukov,Nomura,
Ikeda,Wang,Li,Li2,Ono,Cvetkevic}  

In particular, the present authors with Tanaka and Kontani have constructed a 
minimal model, which has turned out to be five-band, 
for LaFeAsO based on first principles 
calculation, and investigated spin-fluctuation mediated
superconductivity with 
random-phase approximation (RPA).\cite{Kuroki1st,AritaLT,NJP,Physica} 
In that study it was pointed out that, along with the 
sign-reversing $s$-wave, a 
$d$-wave pairing can also be a candidate depending on 
the band filling.  Our five-band model has subsequently been adopted 
in various studies, among which are 
a perturbation study by Nomura\cite{Nomura},
a fluctuation exchange study by Ikeda\cite{Ikeda}, and a functional 
renormalization group study by Wang {\it et al.}\cite{Wang}
An analysis on the normal state spin dynamics has also been 
performed using our five-band model,\cite{Kariyado} where good agreement with 
inelastic neutron scattering experiments\cite{Ishikado,Matan} 
has been obtained.
On the other hand, Graser {\it et al.}\cite{Graser,Maier3} recently 
applied RPA to a five-band model that is based on a band structure 
obtained by Cao {\it et al.}\cite{Cao}  The study finds that 
a sign-reversing $s$-wave that has nodes intersecting the Fermi surface
closely competes with $d$-wave pairing (see Fig.\ref{fig4}). 
It has further been proposed in ref.\onlinecite{Mishra} that 
this nodal $s$-wave pairing is intrinsic to 
the iron pnictide superconductors, while 
a full gap can occur as a consequence of the presence of impurities.
It is also worth noting that 
a competition or mixture of sign-reversing $s$ and $d$ pairings 
have also been discussed on the basis of a 16 band $d$-$p$ model\cite{Ono} 
and a two-orbital exchange coupling 
model ($J_1$-$J_2$ model).\cite{JPHu}
It is the purpose of the present paper to explore systematically 
the material- and structure-dependences on the strength and 
gap symmetry of superconductivity in terms of the five-band 
model.  

Experimentally, the fully-gapped, sign-reversing $s$-wave scenario is 
consistent with 
a number of measurements on arsenides, such 
as angle-resolved photoemission spectroscopy (ARPES),\cite{Ding,Sato} 
penetration depth measurements,\cite{Matsuda} and muon 
spin relaxation ($\mu$SR)\cite{Luetkens,Uemura,Goko,Kadono}, 
which suggest that the gap 
is open on the whole Fermi surface, although the magnitude of the gap 
may vary along the surface. The fully-gapped, sign-reversing $s$ 
is also consistent with 
some neutron scattering results\cite{Christianson,Lumsden} that observe 
a resonance peak predicted theoretically.\cite{Maier,Maier2,Onariresonance}
On the other hand, 
the weak effect of nonmagnetic impurities such as Co on $T_c$ 
\cite{SatoImp} or even the appearance of superconductivity 
upon Co doping\cite{Sefat2,Sefat1,HosonoCobalt} has  
cast doubt on the sign-reversing 
gap, but some theoretical studies\cite{Parker,Senga,Senga2} 
have shown that these  
experiments in fact do not necessarily contradict with the $s\pm$. 
In particular, Senga and Kontani showed that the 
effect of the inter-pocket scattering due to 
nonmagnetic impurities becomes irrelevant in the unitarity limit.
\cite{Senga,Senga2}

One interesting and important feature in the iron pnictides is 
the unusually strong dependence of $T_c$ on materials, 
which ranges from $\simeq 5$K in LaFePO\cite{KamiharaP} 
to 55K in SmFeAsO\cite{Ren} even within the same group of 
elements.  More systematically, Lee {\it et al.}\cite{Lee} 
pointed out that we can parametrize the value of $T_c$ 
in terms of the Fe-Pn-Fe (Pn: pnictogen) bond angle $\alpha$, 
where $T_c$ seems to have a peak around 
the bond angle ($\simeq 109$) at which the pnictogens form a regular 
tetrahedron, while $T_c$ is low for materials 
with large $\alpha$ such as LaFePO. The importance of the 
bond angle has also been pointed out by Zhao {\it et al.}\cite{Zhao}  
On the other hand, it has also been 
shown by Miyazawa {\it et al.} that the chemical trend 
for LnFeAsO has the maximum $T_c$ increasing with 
the decreasing lattice constant $a$ for 
Ln=La $\rightarrow$ Nd, but nearly constant for Ln=Nd $\rightarrow$ 
Dy.\cite{Miyazawa}   
Pressure effects have also been experimentally elaborated. 
For LaFeAsO, $T_c$ first increases with 
pressure, but then decreases when the pressure becomes too 
large,\cite{Takahashi,Fujiwara,Okada,Takeshita2} which is contrasted with 
materials having $T_c > 50$ K at ambient pressure such as NdFeAsO 
for which $T_c$ monotonically and rapidly decreases with 
pressure.\cite{Takeshita}  All these experimental 
results indicate that 
$T_c$ is unusually sensitive to the lattice structure in the iron-based 
compounds.

If we move on to the symmetry of the superconducting gap, 
we have various pieces of experimental evidence for 
strong material dependence as well: while a number of 
experiments on arsenides suggests that the gap is fully open on the 
Fermi surface as mentioned, a recent penetration depth measurement 
on LaFePO shows that there are nodes in the 
superconducting gap.\cite{Fletcher,Hicks}  
Arsenides and LaFePO also exhibit sharp contrast in 
nuclear magnetic resonance (NMR) experiments. In LaFeAsO, 
some experiments show that the NMR relaxation rate 
$1/T_1$ has no coherence peaks, 
and decays as $T^3$ below $T_c$,
\cite{Ishida,Mukuda,Imai,Matano,Kotegawa,Fukazawa} 
while a recent experiment by Kobayashi {\it et al.}
\cite{KobayashiSato} indicates a more rapid decay.
In LaFePO, $1/T_1$ below $T_c$ 
is shown to decay even more slowly than above $T_c$.\cite{Nakai}
These results strongly suggest that even the symmetry of the 
superconducting gap can be unusually sensitive to the lattice structure.

So the crucial theoretical question is how we can 
understand these sensitivities.  In analyzing the 
structure dependence, there is one factor to which 
we have to pay attention.  
Previous theoretical studies have shown that the position of the 
pnictogen with respect to the Fe plane affects  
the band structure, in particular the character of the bands that lie 
close to the Fermi level near the $\Gamma$ point (in the folded 
Brillouin zone) as well as the 
band width (see Fig.\ref{fig7}).\cite{SinghDu,GeorgesArita,Lebegue}  
Local spin-density approximation studies have shown that 
the tendency towards magnetism 
becomes stronger when the pnictogen lies 
farther 
from the Fe plane, which is expected to enhance superconductivity if 
the pairing is mediated by spin fluctuations.\cite{SinghSb}

\begin{figure}[h]
\begin{center}
\includegraphics[width=8.0cm,clip]{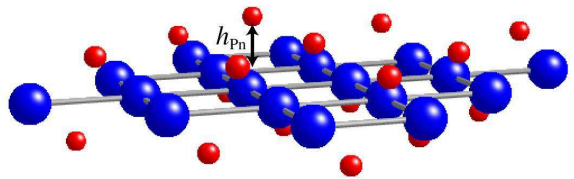}
\caption{(Color online) 
Lattice structure of one Fe-Pn layer, with the pnictogen height 
indicated.
\label{fig1}}
\end{center}
\end{figure}

Given this background, in the present study we investigate the 
lattice structure dependence of the 
spin-fluctuation mediated superconductivity, where we construct 
five-band models for 
several virtual lattice structures of LaFeAsO 
as well as actual materials such as NdFeAsO and LaFePO, and apply 
RPA to solve the Eliashberg equation.
We shall show that 
the position of the pnictogen is indeed the key factor that determines 
both $T_c$ {\it and} the form of the superconducting gap, namely, the 
``pnictogen height"   
above the Fe plane (Fig.\ref{fig1})
can act as a {\it switch between a high-$T_c$, fully-gapped, 
sign-reversing $s$-wave and a 
low-$T_c$, nodal ($s$- or $d$-wave) pairings}.  
We also show that the reduction in the lattice constant is 
generally unfavorable for superconductivity.  Combining these results 
for the lattice structure dependence, 
we then 
obtain a generic ``phase diagram'' against the pnictogen height and 
the lattice constants.  Based on the phase diagram, we argue that 
the systematic dependence of $T_c$ against the bond angle found in 
ref.\onlinecite{Lee} can be accounted for as a 
combined effect of the pnictogen height and the lattice constants.  
In order to get higher $T_c$, 
we propose to seek for materials that have high 
position of the pnictogen and large lattice constants simultaneously.

\begin{figure}[h]
\begin{center}
\includegraphics[width=8.0cm,clip]{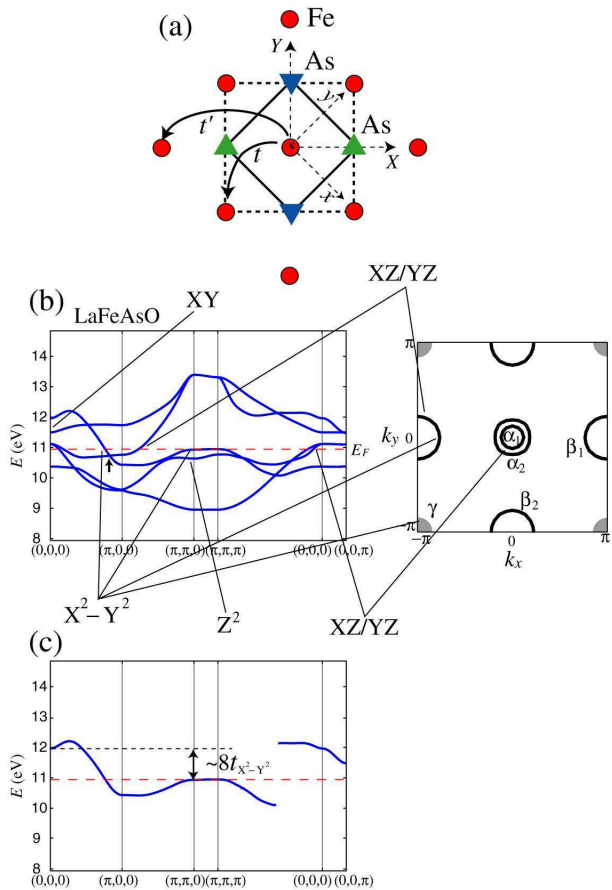}
\caption{(Color online) 
(a) The original (dashed lines) and 
reduced (solid) unit cells with $\bullet$ (Fe), $\nabla$ (As 
below the Fe plane) and $\triangle$ (above Fe). 
(b) The band structure (left) of the five-band model for LaFeAsO, and the 
Fermi surface (right) at $k_z=0$ for $n=6.1$. The main orbital 
characters of some portions of the bands and the Fermi surface 
are indicated.  
The dashed horizontal line in the band structure indicates the Fermi level 
for $n=6.1$. The short arrow in the band structure indicates the 
position of the Dirac cone closest to the Fermi level.  
The gray areas in the Fermi surface 
around the zone corners represent the $\gamma$ Fermi surface.  
(c) The portion of the band that has mainly the $d_{X^2-Y^2}$ orbital 
character.
\label{fig2}}
\end{center}
\end{figure}

\section{The band structure and the Fermi surface}

LaFeAsO has a tetragonal layered structure, where Fe atoms form a 
square lattice in each layer, which is 
sandwiched by As atoms (Figs.\ref{fig1},\ref{fig2}(a)). 
Due to the tetrahedral coordination of As, 
there are two Fe atoms per unit cell.  
The experimentally determined lattice constants 
are $a=4.036$\AA \ and $c=8.739$\AA, with 
two internal coordinates $z_{\rm La}=0.142$ and $z_{\rm As}=0.6512$.
\cite{Hosono}
We have obtained the band structure (Fig.\ref{fig2}(b) ) with the 
local density approximation with a plane-wave basis\cite{pwscf}.  
We then construct the maximally-localized 
Wannier functions (MLWFs)\cite{MaxLoc}. 
These MLWFs, centered at the two Fe sites in the unit cell, 
have five orbital symmetries ($d_{3Z^2-R^2}$, 
$d_{XZ}$, $d_{YZ}$, $d_{X^2-Y^2}$, and 
$d_{XY}$, where $X, Y, Z$ refer to those for the unit cell 
with two Fe sites as shown in Fig.\ref{fig2}(a)).  
The two Wannier orbitals in 
each unit cell are equivalent in that each Fe atom has the same 
local arrangement of surrounding atoms.  
We can then take a unit cell that 
contains only one orbital (for each orbital symmetry) by 
unfolding the Brillouin zone,
and we end up with an effective five-band model on a 
square lattice, where 
$x$ and $y$ axes are rotated by 
45 degrees from $X$-$Y$, 
to which we refer for all the wave vectors hereafter. 
We define the band filling $n$ as the number of electrons/number of sites
(e.g., $n=10$ for a full filling). 
The doping level $x$ 
in LaFeAsO$_{1-x}$F$_x$ is related to the band filling as $n=6+x$.

The five bands are heavily entangled as shown in Fig.\ref{fig2}(b) 
reflecting the strong hybridization 
of the five $3d$ orbitals, which is physically due to the tetrahedral 
coordination of As atoms around Fe.  
Hence we conclude that the minimal electronic model 
requires all the five bands.\cite{AritaLT} 
In Fig.\ref{fig2}(b), 
the Fermi surface at $k_z=0$ for $n=6.1$ (corresponding to $x=0.1$) 
is shown in the unfolded Brillouin zone.  
The Fermi surface consists of four pieces (pockets in 2D):   
two concentric hole pockets (denoted as $\alpha_1$, $\alpha_2$) 
around $(k_x, k_y)=(0,0)$, two electron pockets 
around $(\pi,0)$ $(\beta_1)$ or $(0,\pi)$ $(\beta_2)$, respectively. 
Besides these pieces of the Fermi surface, there is a portion of the band 
around $(\pi,\pi)$ that touches the $E_F$ at $n=6.1$, so that 
this portion acts as a ``quasi Fermi surface" (which we call $\gamma)$.  
As for the orbital character, $\alpha$ and some portions of $\beta$ near 
the Brillouin zone edge have mainly $d_{XZ}$ and $d_{YZ}$ character, 
while the portions of 
$\beta$ away from the Brillouin zone edge and $\gamma$ have 
mainly $d_{X^2-Y^2}$ orbital character.  
An interesting feature in the band structure 
is the presence of Dirac cones, i.e., 
places where the upper and the lower bands make a conical contact.
\cite{NJP,Fukuyama}
The ones closest to the Fermi level correspond to the crossing points of 
the $d_{X^2-Y^2}$ and the $d_{XZ}/d_{YZ}$ bands below the 
$\beta$ Fermi surface. 

\section{Many-body Hamiltonian and random-phase approximation}

For the many body part of the Hamiltonian, 
we consider the standard interaction terms that comprise 
the intra-orbital Coulomb $U$, the inter-orbital 
Coulomb $U'$, the Hund's coupling $J$, and the pair-hopping $J'$. 
The many body Hamiltonian then reads
\begin{eqnarray}
H &=& \sum_i\sum_\mu\sum_{\sigma}\varepsilon_\mu n_{i\mu\sigma}
 + 
\sum_{ij}\sum_{\mu\nu}\sum_{\sigma}t_{ij}^{\mu\nu}
c_{i\mu\sigma}^{\dagger} c_{j\nu\sigma}\nonumber\\
&+&\sum_i\left( U\sum_\mu n_{i\mu\uparrow} n_{i\mu\downarrow}
+U'\sum_{\mu > \nu}\sum_{\sigma,\sigma'} n_{i\mu\sigma} n_{i\nu\sigma'}
\right.\nonumber\\
&&\left.-J\sum_{\mu\neq\nu}\Vec{S}_{i\mu}\cdot\Vec{S}_{i\nu}
+J'\sum_{\mu\neq\nu} c_{i\mu\uparrow}^\dagger c_{i\mu\downarrow}^\dagger
c_{i\nu\downarrow}c_{i\nu\uparrow}
\right), 
\end{eqnarray}
where $i,j$ denote the sites and $\mu,\nu$ the (five d) orbitals, 
and $t_{ij}^{\mu\nu}$ is the  obtained in the 
previous section.
The orbitals $d_{3Z^2-R^2}$, 
$d_{XZ}$, $d_{YZ}$, $d_{X^2-Y^2}$, and 
$d_{XY}$ are labeled as $\nu=1,2,3,4,$ and 5, respectively.
As for the electron-electron interactions, there have been 
theoretical studies that estimate the parameter values.  
Some studies give $U=2.2-3.3$ and $J=0.3-0.6$\cite{Nakamura,Miyake} 
in units of eV, 
while others have $U\sim J$.\cite{Anisimov}  
Here we assume that $U>J$  and take the values 
$U=1.2$, $U'=0.9$, and $J=J'=0.15$.
We also examine orbital-dependent interactions 
as introduced in section\ref{nakamura}. 
We have taken the values somewhat smaller than 
those obtained in ref.\onlinecite{Nakamura,Miyake} because 
the self-energy correction is not taken into account in the 
present RPA calculation, so that small interaction parameters 
are needed to avoid magnetic ordering at high temperatures.

Having constructed the model, we move on to the five-band RPA calculation, 
where the modification of the band structure due to 
the self-energy correction is not taken into account. 
Multiorbital RPA is described in e.g.
ref.\onlinecite{Yada,Takimoto}. In the present case, 
Green's function $G_{lm}(k)$ $(k \equiv (\Vec{k},i\omega_n))$
is a $5\times 5$ matrix. The irreducible susceptibility matrix 
\begin{equation} 
\chi^0_{l_1,l_2,l_3,l_4}(q) =\sum_k G_{l_1l_3}(k+q)G_{l_4l_2}(k)
\end{equation}
$(l_i = 1,...,5)$ has $5^4$ components, and 
the spin and the charge (orbital) susceptibility matrices are obtained 
from matrix equations, 
\begin{equation}
\hat{\chi}_s(q)=\frac{\hat{\chi}^0(q)}{1-\hat{S}\hat{\chi}^0(q)} ,
\end{equation}
\begin{equation}
\hat{\chi}_c(q)=\frac{\hat{\chi}^0(q)}{1+\hat{C}\hat{\chi}^0(q)} ,
\end{equation}
where 
\begin{equation}
S_{l_1l_2,l_3l_4}
=\left\{\begin{array}{cc}
U, &\;\; l_1=l_2=l_3=l_4\\ 
U',&\;\; l_1=l_3\neq l_2=l_4\\
J,&\;\; l_1=l_2\neq l_3=l_4\\
J',&\;\; l_1=l_4\neq l_2=l_3 ,
\end{array}  \right.
\end{equation}

\begin{equation}
C_{l_1l_2,l_3l_4}
=\left\{\begin{array}{cc}
 U &\;\; l_1=l_2=l_3=l_4\\ 
-U'+J & \;\; l_1=l_3\neq l_2=l_4\\
2U'-J,&\;\; l_1=l_2\neq l_3=l_4\\
J'& \;\; l_1=l_4\neq l_2=l_3 .
\end{array}  \right.
\end{equation}
We denote the largest eigenvalue of the spin (charge) 
susceptibility matrix for $i \omega_n=0$ 
as $\chi_s(\Vec{k}) (\chi_c(\Vec{k}))$. 

The Green's function and 
the effective singlet pairing interaction, 
\begin{equation}
\hat{V}^s(q)=\frac{3}{2}\hat{S}\hat{\chi}_s(q)\hat{S}
-\frac{1}{2}\hat{C}\hat{\chi}_c(q)\hat{C}+\frac{1}{2}(\hat{S}+\hat{C}),
\end{equation}
are plugged into the linearized Eliashberg equation, 
\begin{eqnarray}
\lambda \phi_{l_1l_4}(k)&=&-\frac{T}{N}\sum_q
\sum_{l_2l_3l_5l_6}V_{l_1 l_2 l_3 l_4}(q)\nonumber\\
&\times& G_{l_2l_5}(k-q)\phi_{l_5l_6}(k-q)
G_{l_3l_6}(q-k).
\end{eqnarray}
The $5\times 5$ matrix gap function $\phi_{lm}$ 
in the orbital representation 
along with the associated eigenvalue $\lambda$ is obtained by 
solving this equation. 
The gap function can be transformed into the band representation 
with a unitary transformation. 
The calculation is performed at $T=0.02$ eV taking a three 
dimensional $k$-point mesh of $32\times 32\times 4$ and  512 
Matsubara frequencies. All the results for the spin susceptibility and the 
superconducting gap will be presented for the 
lowest Matsubara frequency and at $k_z=0$ or $q_z=0$.
The eigenvalue of the Eliashberg equation $\lambda$ at the fixed 
temperature of 0.02eV will 
be adopted as a measure of the strength of the superconducting 
instability, since directly obtaining $T_c$, especially for low $T_c$ 
systems, requires more $k$-point meshes and Matsubara frequencies.

\section{Orbital-dependent nesting and the pairing symmetry competition}

\begin{figure}[h]
\begin{center}
\includegraphics[width=8.0cm,clip]{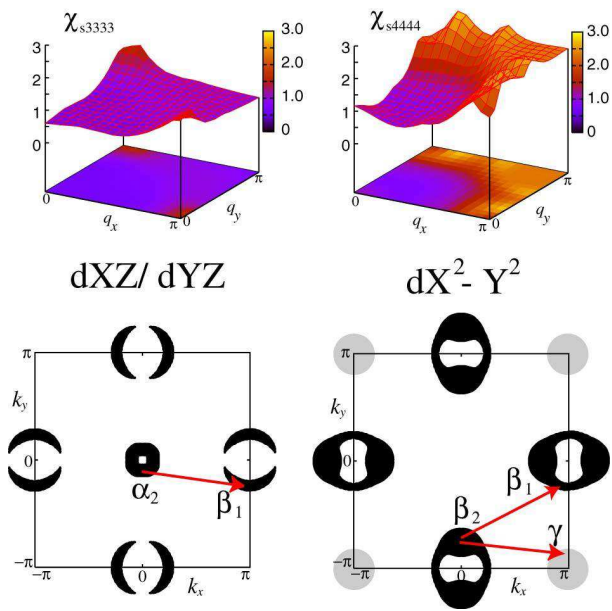}
\caption{(Color online) 
Top panels : Diagonal components, $\chi_{s3333}$ and 
$\chi_{s4444}$, of the 
spin susceptibility matrix in the orbital representation 
($3: YZ, 4: X^2-Y^2$) for the five-band model of LaFeAsO with $n=6.1$.  
Bottom: Nesting is shown for the Fermi surface 
for orbitals $XZ, YZ$ (left)
and $X^2-Y^2$ (right). Here the thickness of the Fermi surface 
represents the strength of the respective orbital character. 
\label{fig3}}
\end{center}
\end{figure}

Let us first look in Fig.\ref{fig3} at the result for 
the (orbital-diagonal components of) spin susceptibility, 
$\chi_{s3333}$ and $\chi_{s4444}$, which are 
the two largest components.  
$\chi_{s3333}$ has peaks solely around $(\pi,0)$ and $(0,\pi)$, 
which reflects the nesting between $d_{XZ}, d_{YZ}$ portions of 
$\alpha$ and $\beta$ pockets as shown in the lower panel of 
Fig.\ref{fig3}, where the thickness of the Fermi surface represents the 
strength of the $d_{X^2-Y^2}$ or $d_{XZ}/d_{YZ}$ characters.  
On the other hand, $\chi_{s4444}$ has peaks around $(\pi,0), (0,\pi)$ 
and $(\pi,\pi/2), (\pi/2,\pi)$. The former is due to the nesting 
between the $\gamma$ pocket and the $d_{X^2-Y^2}$ 
portion of the $\beta$ pocket, while the latter originates 
from the nesting between the $d_{X^2-Y^2}$ 
portion of the $\beta_1$ and $\beta_2$.
\cite{Kuroki1st,Physica,Graser,Maier3,commentbeta}

\begin{figure}[h]
\begin{center}
\includegraphics[width=4.0cm,clip]{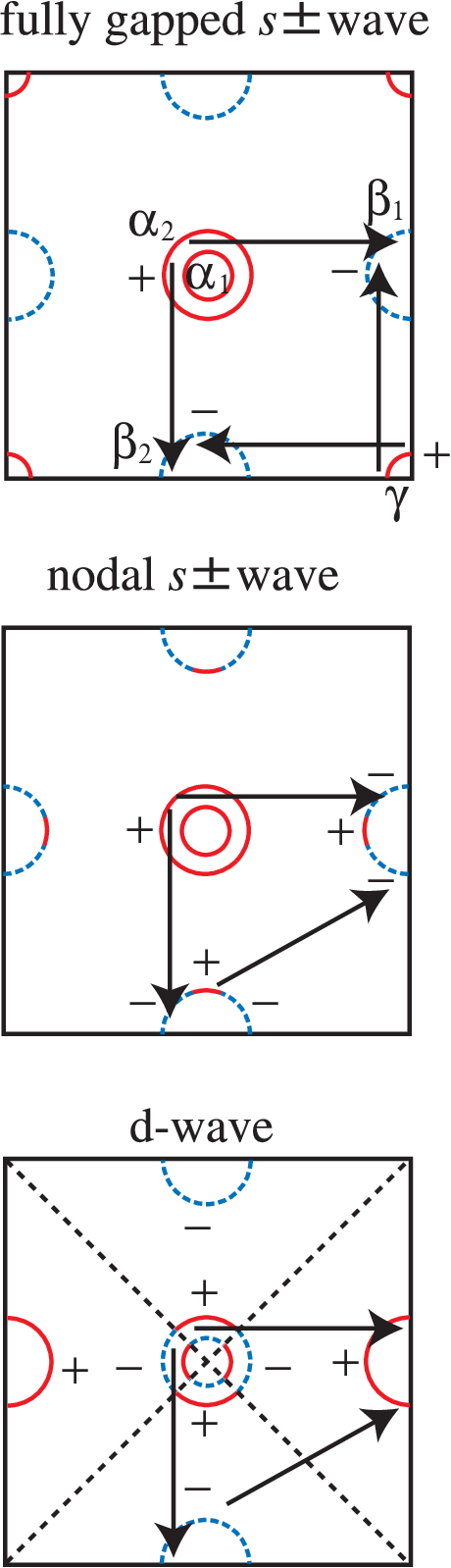}
\caption{(Color online) 
The fully-gapped $s\pm$ wave (top panel), the nodal 
$s\pm$ wave (middle), and the $d$-wave gap (bottom) 
are schematically shown. The solid red (dashed blue) curves 
represent positive (negative) sign of the gap. The arrows 
indicate the dominating nesting vectors.
$\gamma$ Fermi surface is present when the pnictogen height is large.
\label{fig4}}
\end{center}
\end{figure}

\begin{figure}[h]
\begin{center}
\includegraphics[width=8.0cm,clip]{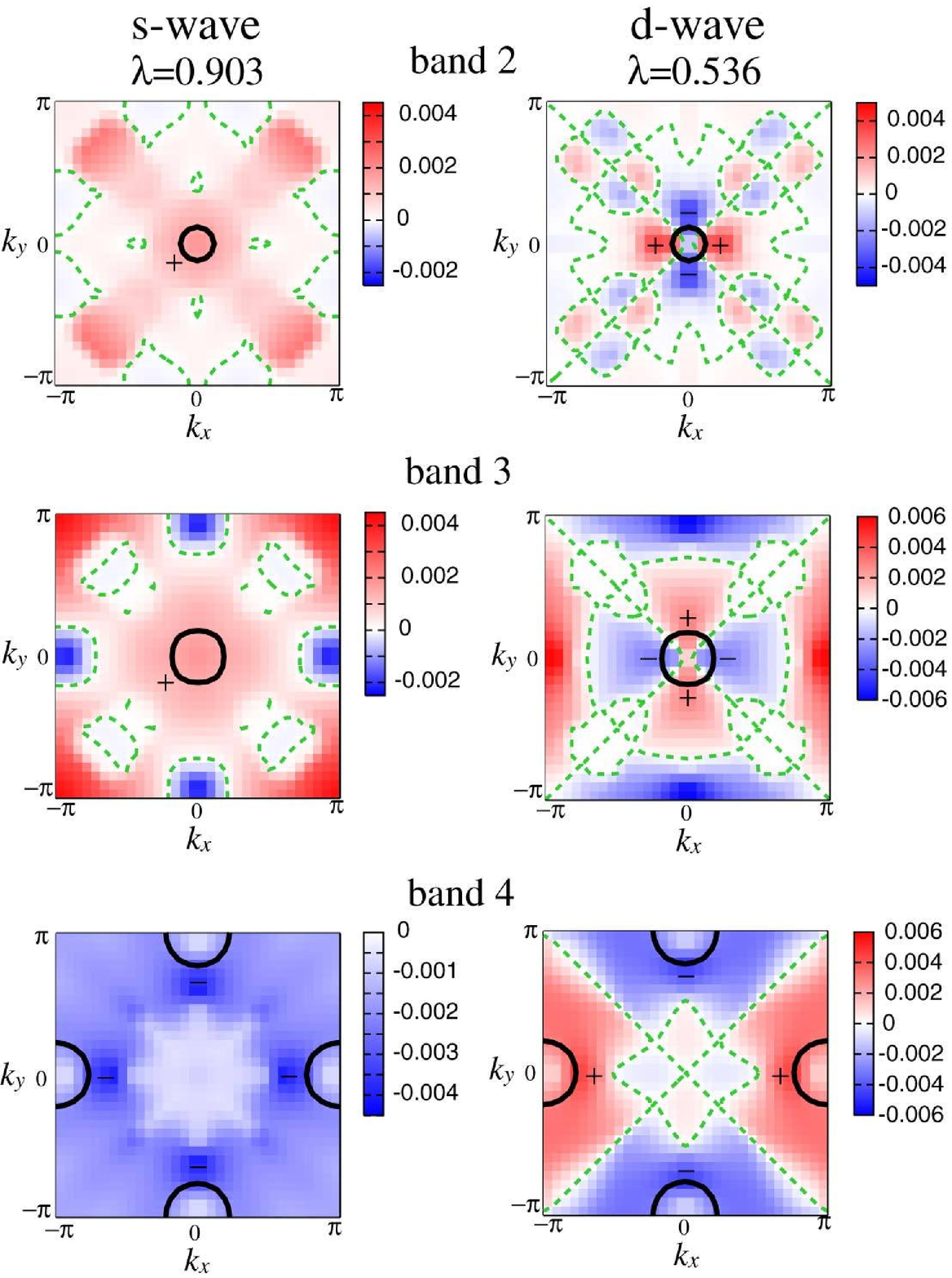}
\caption{(Color online) 
The $s$-wave (left panels) and $d$-wave (right) 
gap functions for the 2nd to 4th bands from top to bottom 
in the band representation 
for the five-band model of LaFeAsO with $n=6.1$. 
Solid lines represent the Fermi surface, and 
green dashed lines the nodes in the gap.
\label{fig5}}
\end{center}
\end{figure}

The superconducting gap 
should be determined by the cooperation or competition between the multiple 
nestings mentioned above.  
Specifically, the $\alpha$-$\beta$ and $\gamma$-$\beta$ 
nestings tend to favor the fully-gapped, sign-reversing $s$-wave, in 
which the gap changes sign between $\alpha$ and $\beta$ but has 
a constant sign on each pocket as shown in 
Fig.\ref{fig4}.\cite{Mazin}  
On the other hand, $\beta_1$-$\beta_2$ nesting tends to change the 
sign of the gap between these pockets, which can result in 
either $d$-wave or an $s$-wave pairing with nodes on the 
$\beta$ Fermi surface, as shown schematically in Fig.\ref{fig4}
\cite{Kuroki1st,Graser,errata}.
For the band structure of LaFeAsO (obtained by using the experimentally 
determined lattice structure), the sign-reversing 
$s$-wave with no nodes intersecting the Fermi surface 
dominates for the present set of parameter values with $n=6.1$ as shown in 
Fig.\ref{fig5}.\cite{errata} The eigenvalue of the Eliashberg equation 
at $T=0.02$eV is $\lambda=0.90$ for $s$-wave,
against $\lambda=0.54$ for $d$-wave.
\begin{figure}[h]
\begin{center}
\includegraphics[width=8.0cm,clip]{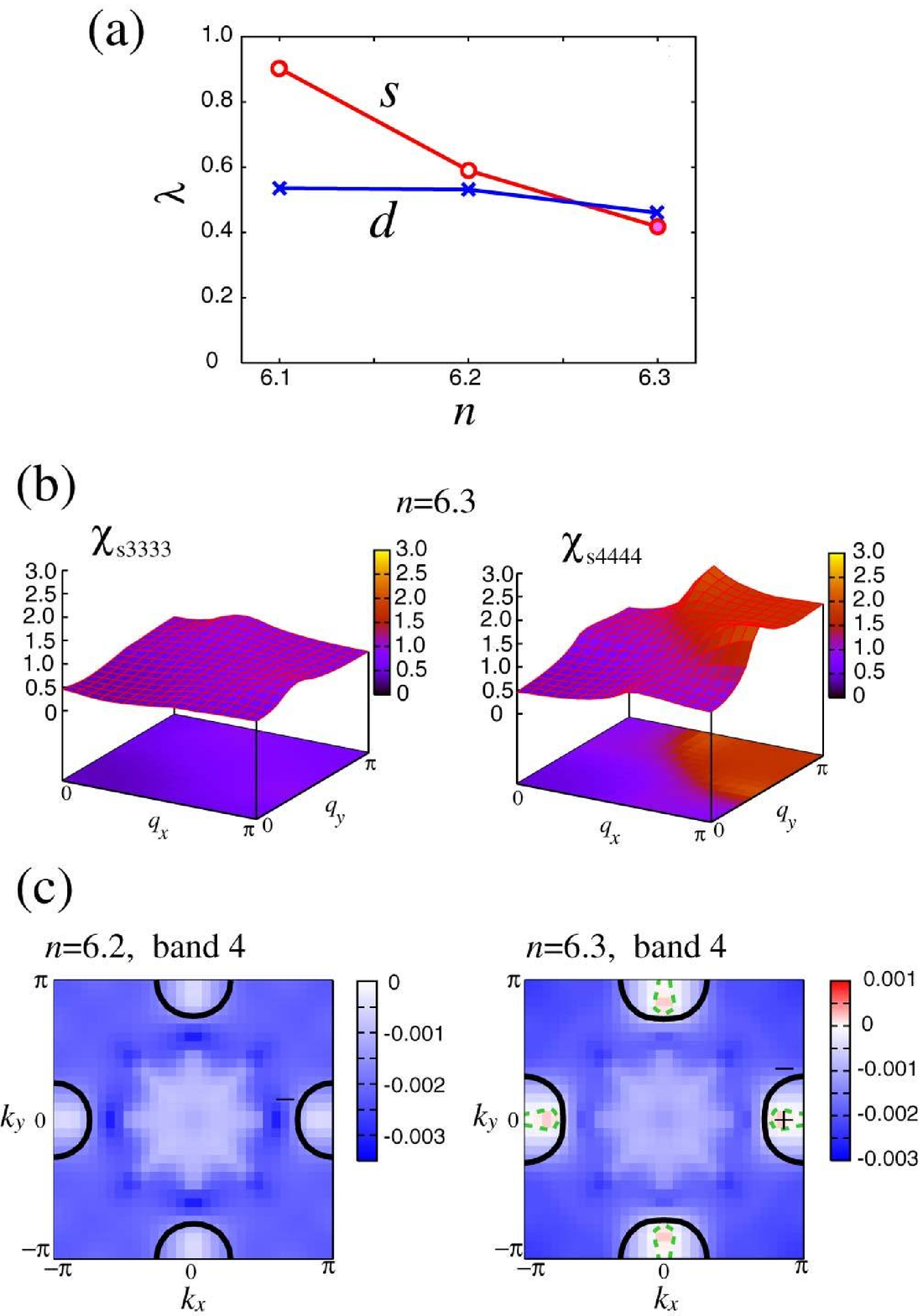}
\caption{(Color online) 
(a) Eigenvalues of the 
Eliashberg equation for s-wave and d-wave, respectively, 
for the five-band model of 
LaFeAsO plotted against the band filling $n$. 
The light red (or gray) symbol for the $s$-wave 
at $n=6.3$ indicates that the gap is nearly nodal.
(b) $\chi_{s3333}$ and $\chi_{s4444}$ for $n=6.3$.   
(c) The $s$-wave 
gap function for band 4 with $n=6.2$(left) and $n=6.3$(right).
\label{fig6}}
\end{center}
\end{figure}

As for the band filling dependence, we plot the 
eigenvalue of the Eliashberg equation of $s$- and $d-$wave pairings 
in Fig.\ref{fig6}(a). 
We can see for the band structure of LaFeAsO with the 
present set of interaction values that 
the sign-reversing $s$-wave pairing with a full gap for each 
pocket dominates 
for the band filling $n\leq 6.2$. For $n\geq 6.3$, 
the $\gamma$ pocket becomes less effective, 
and the $(\pi,0)$ peak in $\chi_{s4444}$ disappears as seen in 
the right panel of Fig.\ref{fig6}(b). 
The $\alpha$ pocket becomes less effective as well, 
and the $(\pi,0)$ peak in $\chi_{s3333}$ becomes small.
Thus in this region, 
$d$-wave pairing begins to dominate, and the subdominant $s$-wave 
gap has nodes 
almost touching the $\beta$ as seen in the right panel 
of Fig.\ref{fig6}(c).  For small doping levels when the $\gamma$ Fermi surface 
is effective and $s$-wave dominates, the magnitude of the 
$s$-wave gap has maxima 
at the positions along the $\beta$ pocket 
facing the $\Gamma$ point (Fig.\ref{fig5}, lower left), 
but when the doping increases to $n=6.3$, the $s$-wave 
gap has minima at these points. The gap turns out to be nearly 
constant on the $\beta$ Fermi surface (Fig.\ref{fig6}(c), left)
for the band filling $n\simeq 6.2$. In this case, the gap on $\alpha$ 
(not shown) and $\beta$ have nearly the same magnitude.

Note however that the present analysis on the 
band filling dependence does not take account of the doping dependence of the 
band structure itself, which should occur mainly due to the  
change of the As position caused by doping. 
We will come back to this point in section\ref{ndfeaso}, 
taking NdFeAsO as an example.

\section{The effect of the lattice structure}
\subsection{Pnictogen height dependence}

\begin{table*}
\begin{tabular}{c|c|c|c|c|c|r|r|r|r}
 & $a$(\AA) & $c$(\AA)  & $z_{\rm Pn}$ &$h_{\rm Pn}$ (\AA) & $\alpha$ &  $t_{X^2-Y^2}$ &  $t'_{X^2-Y^2}$ & $t_{XZ}$ & $t'_{XZ}$ \\ \hline
 La\cite{Hosono}  & 4.04& 8.74 & 0.6512 & 1.32 &113.6& 0.163  & 0.124  & $-0.210$  & 0.329 \\
$h_{\rm As}=1.38$\AA & 4.04& 8.74 & 0.6580 &  1.38 &111.2  & 0.132  & 0.113  & $-0.191$  & 0.309 \\
$h_{\rm As}=1.14$\AA & 4.04& 8.74 & 0.6304 & 1.14 &121.1 & 0.261  & 0.153  & $-0.240$  & 0.364 \\
 $a=3.95$\AA&       3.95& 8.74 & 0.6512 & 1.32 &112.4  & 0.148  & 0.123  & $-0.210$  & 0.346    \\
 $c=8.40$\AA&       4.04& 8.40 & 0.6573 & 1.32 &113.6  & 0.174  & 0.132  & $-0.209$  & 0.327  \\
 Nd\cite{Lee} &    3.94& 8.51 & 0.6624 & 1.38 &109.9 & 0.135  & 0.123  & $-0.202$  & 0.332   \\ 
 Nd-p\cite{Kumai} & 3.92& 8.37 & 0.6584 & 1.33 &111.9 & 0.172  & 0.138  & $-0.217$  & 0.350  \\
 Nd-ud\cite{Lee}  &  3.97& 8.57 & 0.6571 & 1.35 &111.7 & 0.156  & 0.129  & $-0.213$  & 0.341 \\
 P\cite{KamiharaP} &  3.96&  8.51 & 0.6339 & 1.14 & 120.2 & 0.253  & 0.156  & $-0.234$  & 0.377 \\
\end{tabular}
\caption{(Color online) 
Materials and lattice structures considered in the present study, 
and the nearest and second-nearest neighbor hopping integrals (in eV) 
in the corresponding tight-binding models. 
Shorthands are: La (LaFeAsO), Nd (optimally doped NdFeAsO$_{1-y}$), 
Nd-p (NdFeAsO$_{1-y}$ under the pressure of 3.8GPa), Nd-ud 
(underdoped NdFeAsO$_{1-y}$), and P (LaFePO).  
The cases with $a=3.95$\AA \ , $c=8.40$\AA \ , 
$h_{\rm As}=1.38$\AA \, or 1.14\AA \ correspond to 
virtual structures of 
LaFeAsO. Note that $(z_{\rm Pn}-0.5)\times c=h_{\rm Pn}$.
\label{table1}
}
\end{table*} 

\begin{figure}[h]
\begin{center}
\includegraphics[width=8.0cm,clip]{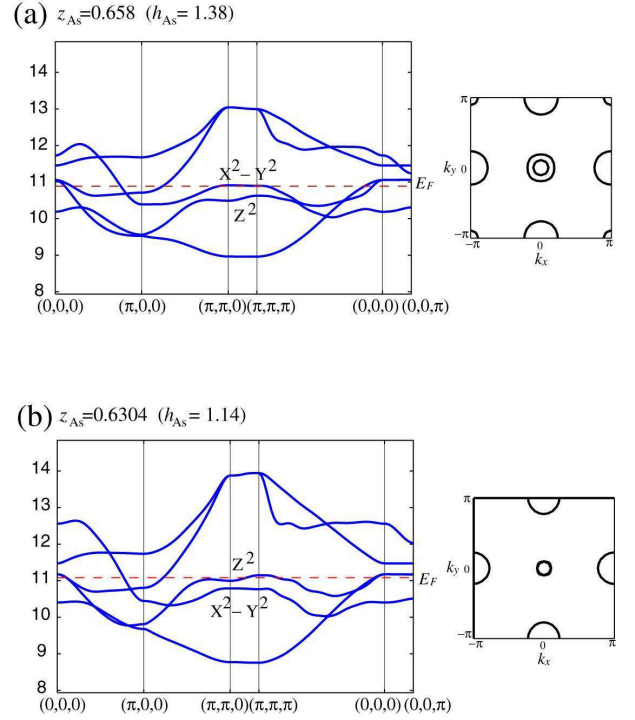}
\caption{(Color online) 
Band structure in the five-band model 
for $z_{\rm As}=0.658$ 
(a) and $z_{\rm As}=0.6304$ (b). The lattice constants are kept at the 
original values of LaFeAsO. $X^2-Y^2$ and $Z^2$ denote the main characters of 
the bands around $(\pi,\pi,k_z)$. Dashed lines represent the Fermi 
energy, and the Fermi surface at $k_z=0$ 
for $n=6.1$ is shown on the right panels.
\label{fig7}}
\end{center}
\end{figure}

We now investigate the effect of the ``pnictogen height $h_{\rm Pn}$ '', 
namely, the distance between a pnictogen atom and the Fe layer, 
$(z_{\rm Pn}-0.5)\times c$, 
where $z_{\rm Pn}$ is the internal coordinate of the 
pnictogen atom and $c$ the c-axis lattice constant. 
As shown in previous studies,\cite{SinghDu,GeorgesArita,Lebegue}  
$z_{\rm Pn}$ controls the relative position of 
$d_{X^2-Y^2}(=d_{xy})$ and $d_{Z^2}$ bands near the 
$\Gamma$ point of the folded (original) Brillouin zone.
In the unfolded Brillouin zone, these bands appear near $(\pi,\pi)$.  
In Fig.\ref{fig7}, we show the band structure in the unfolded Brillouin 
zone for virtual lattice structure with 
$z_{\rm As}=0.658$ and $0.6304$ with   
the lattice constants and $z_{\rm La}$ fixed at the original 
values for LaFeAsO.  
For the original $z_{\rm As}=0.651$, the As height is $h_{\rm As}=1.32$\AA. 
For $z_{\rm As}=0.658$,  $h_{\rm As}$ increases to $1.38$\AA , which is the 
same as in the optimally doped NdFeAsO, 
while $z_{\rm As}=0.6304$ ($h_{\rm As}=1.14$\AA) 
corresponds to the height of P in LaFePO.  
We see that the $d_{X^2-Y^2}$ band that forms the 
$\gamma$ Fermi surface around $(\pi,\pi)$ rises as 
$h_{\rm As}$ is increased, while the $d_{Z^2}$ band 
sinks below the Fermi level.

The reason for these shifts in the band positions can be 
understood in terms of the hopping integrals. 
As $h_{\rm As}$ increases, the nearest-neighbor 
hopping for the $d_{X^2-Y^2}$ orbital decreases 
as shown in Table.\ref{table1}. If we approximate the 
$d_{X^2-Y^2}$ portion of the bands by 
\begin{eqnarray}
\varepsilon_{X^2-Y^2}(\Vec{k})&=&-2t_{X^2-Y^2}[\cos(k_x)+\cos(k_y)]\nonumber\\
&&-4t_{X^2-Y^2}'\cos(k_x)\cos(k_y),
\end{eqnarray}
where $t_{X^2-Y^2}'$ stands for the second-nearest neighbor hopping, 
the energy difference between $(0,0)$ and $(\pi,\pi)$ is 
proportional to $t_{X^2-Y^2}$ (Fig.\ref{fig2}(c)), 
so that the reduction in $t_{X^2-Y^2}$ acts to push up the 
$d_{X^2-Y^2}$ band at $(\pi,\pi)$.  
Besides the variation in the $d_{X^2-Y^2}$ hoppings, 
the increase in $h_{\rm As}$ results in an overall 
reduction in the hopping integrals of other orbitals 
because the effective hopping 
path Fe$\rightarrow$ As$\rightarrow$ Fe becomes less effective.

\begin{figure}[h]
\begin{center}
\includegraphics[width=8.0cm,clip]{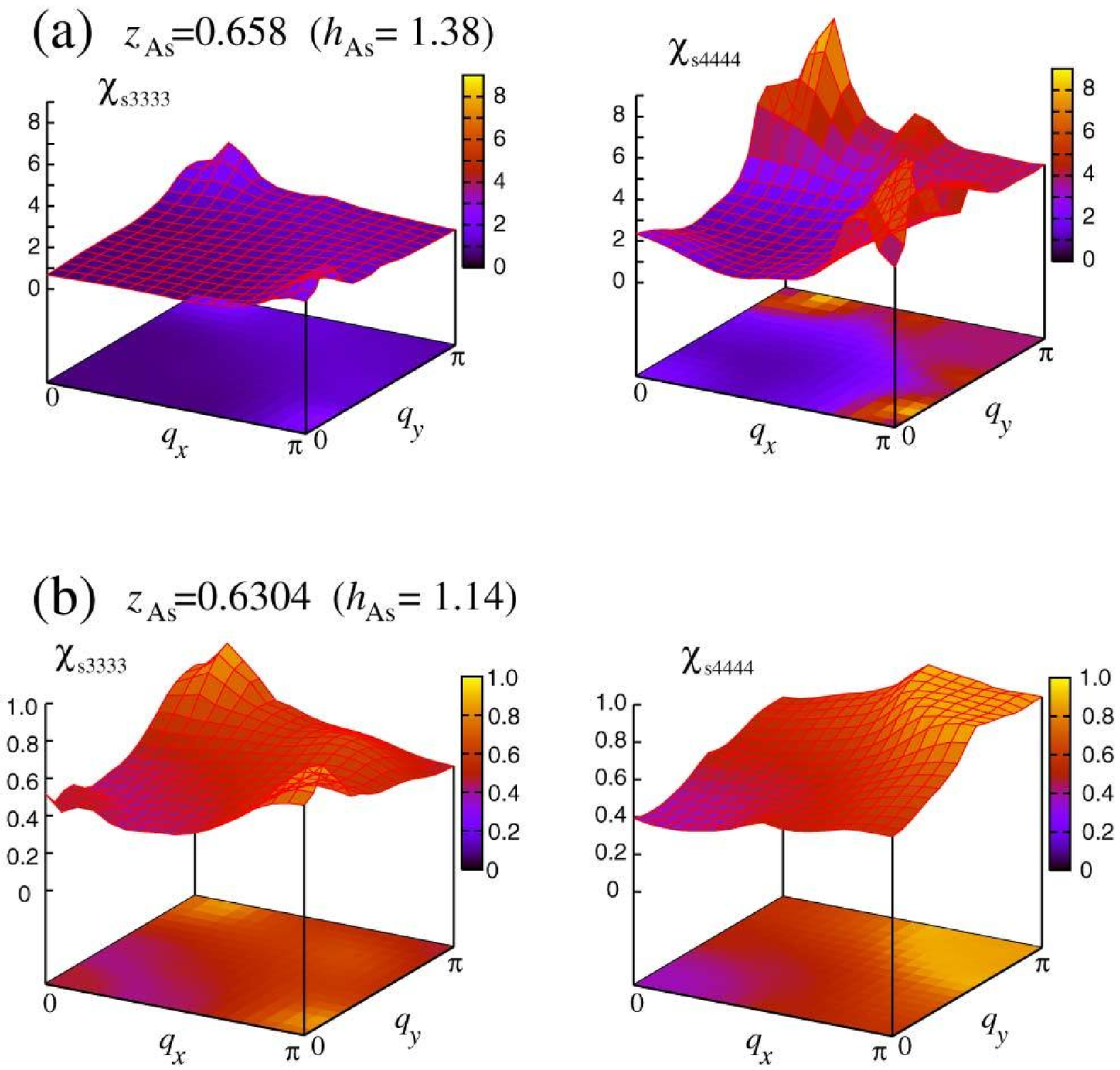}
\caption{(Color online) 
$\chi_{s3333}$ (left panels) and $\chi_{s4444}$ (right) for 
the model with (a) $z_{\rm As}=0.658$ and (b) $z_{\rm As}=0.6304$.
\label{fig8}}
\end{center}
\end{figure}

The effect of varying $h_{\rm As}$  on the spin susceptibility is shown in 
Fig.\ref{fig8}. The $d_{YZ}$ orbital component always has peaks
around $(\pi,0)$,$(0,\pi)$ reflecting the $\alpha$-$\beta$ nesting.
On the other hand, the $d_{X^2-Y^2}$ orbital component of the 
spin susceptibility $\chi_{s4444}$ exhibits a strong variation 
with $h_{\rm As}$:  when $h_{\rm As}$ is large and the 
$\gamma$ pocket is present, $\chi_{s4444}$ (Fig.\ref{fig8}(a), right) 
is strongly peaked at 
$(\pi,0)$, reflecting the $\gamma$-$\beta$ nesting and also the 
strong electron 
correlation due to the overall reduction in the band width.  
However, as $h_{\rm As}$ is reduced, the structure around $(\pi,\pi/2)$,
which arises from the $\beta_1$-$\beta_2$ nesting, dominates 
(Fig.\ref{fig8}(b), right) . 
The effect of reducing $h_{\rm As}$ resembles the effect of electron
doping, but in the case of electron doping, 
not only the effect of the $\gamma$ pocket, but also that of the
$\alpha$ becomes weak, so that $\chi_{s3333}$ is suppressed.

\begin{figure}[h]
\begin{center}
\includegraphics[width=8.0cm,clip]{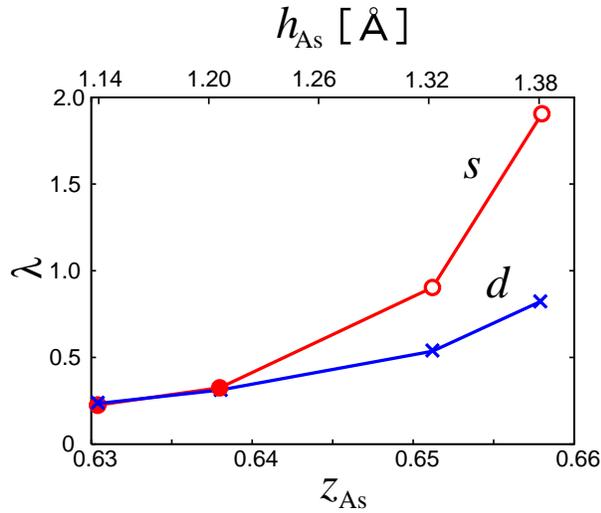}
\caption{(Color online) 
$s$-wave and $d$-wave eigenvalues of the Eliashberg 
equation plotted against $z_{\rm As}$ (lower scale) 
or $h_{\rm As}$ (upper scale) for $n=6.1$. 
The lattice constants are fixed at the original values for LaFeAsO.
For the $s$-wave, the open (solid) circles indicate that the gap is 
nodeless (nodal).
\label{fig9}}
\end{center}
\end{figure}

\begin{figure}[h]
\begin{center}
\includegraphics[width=8.0cm,clip]{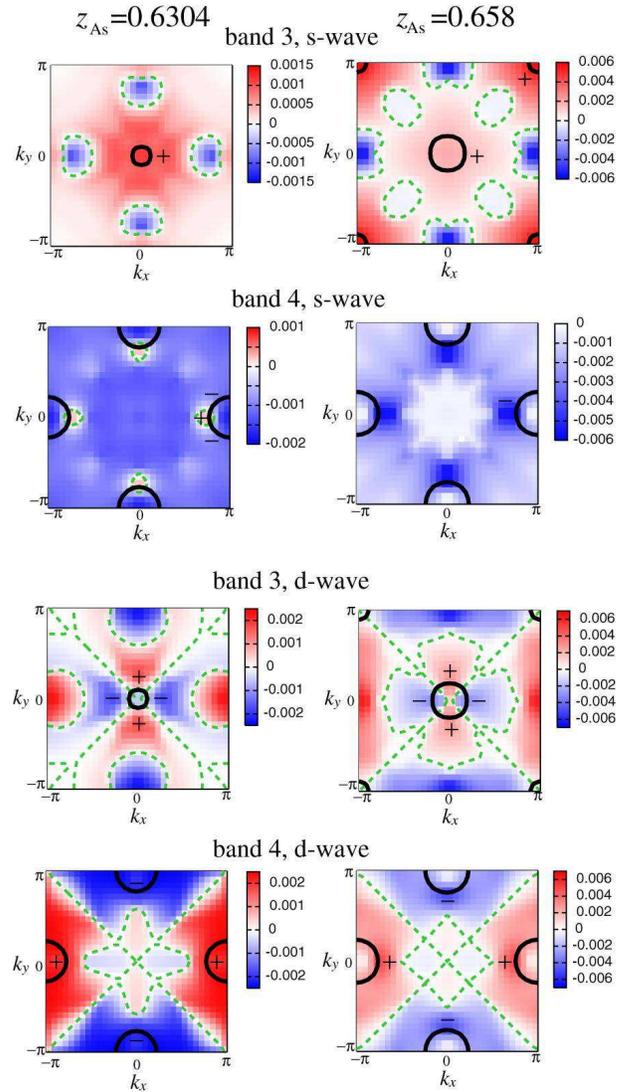}
\caption{(Color online) 
The gap functions for the model with 
$z_{\rm As}=0.6304$ (left panels) or $z_{\rm As}=0.658$ (right).   
From top to bottom: $s$-wave in band 3, $s$-wave in band 4, 
$d$-wave in band 3, $d$-wave in band 4.  
\label{fig10}}
\end{center}
\end{figure}

The effect on the spin susceptibility in turn affects 
superconductivity. When $h_{\rm As}$ is large, 
$\chi_{s4444}$ and $\chi_{s3333}$ 
spin fluctuations near $(\pi,0)$ {\it cooperate} to mediate 
the fully-gapped, sign-reversing $s$-wave superconductivity.
When $h_{\rm As}$ is small, by contrast, 
the $(\pi,\pi/2)$ spin fluctuations begin to 
favor $d$-wave and nodal $s$-wave pairings.  
In Fig.\ref{fig9}, we plot against $z_{\rm As}$ (lower scale) 
or against $h_{\rm As}$ (upper scale) 
the eigenvalue of the Eliashberg equation for
the $s$-wave and $d$-wave pairings, respectively. 
For large $h_{\rm As}$ where
the fully-gapped, sign-reversing $s$-wave 
(Fig.\ref{fig10}, upper right) dominates, 
$\lambda$ is large because the strong spin fluctuations arising from 
$\alpha$-$\beta$ and $\gamma$-$\beta$ nestings cooperate.
This is contrasted with the case of small $h_{\rm As}$, where 
$d$-wave or nodal $s$-wave begin to dominate (Fig.\ref{fig10} left). 
In this region 
$\lambda$ is small because the 
$\gamma$-$\beta$ nesting is no longer effective, or to be 
more precise, its remaining effect {\it competes} with the 
effect of the $\beta_1$-$\beta_2$ nesting.  
It is worth noting that if we adopt $z_{\rm As}=0.638$, 
which is the 
value determined by theoretical structure optimization,\cite{Mazin} 
we have the 
closely competing $d$-wave and nodal $s$-wave pairings.
This is consistent with a recent RPA calculation by Graser {\it et al.},
\cite{Graser} who adopted a band structure determined by a theoretical 
structure optimization.\cite{Cao}
The message of the present analysis, then, is that 
the pnictogen height can act as a ``switch'' between 
the fully-gapped, sign-reversing high-$T_c$ 
$s$-wave and the low-$T_c$ gapless (either $d$-wave or 
nodal $s$-wave) superconductivity.

\subsection{Lattice constant dependence}
We now turn to the effect of the lattice constants. 
We consider virtual lattice structures where 
one of the lattice constants, $a$ or $c$, is varied, 
while the pnictogen height is 
fixed at the original value for LaFeAsO.  
When we reduce the lattice constant $a$, we find that 
the nearest-neighbor hopping $t_{X^2-Y^2}$ decreases, 
probably because the Fe-As-Fe angle is reduced, 
which may cause a suppression of the effective hopping 
via the path Fe$\rightarrow$ As$\rightarrow$ Fe.
Nonetheless, 
most of the other in-plane hopping integrals (including the 
ones not listed in the table) are enhanced as intuitively expected.
On the other hand,  a reduction in the lattice constant $c$ 
is found to mainly enhance the 
in-plane $d_{X^2-Y^2}$ hopping (apart from the 
obvious enhancement of the hopping in the $c$ direction).  
This may be because the As wave function is pushed toward the Fe plane 
for a reduced layer-layer distance, and the hopping between $d_{X^2-Y^2}$ 
orbitals, which is elongated in the direction of the As atom positions,  
is enhanced by this deformation.

\begin{figure}[h]
\begin{center}
\includegraphics[width=8.0cm,clip]{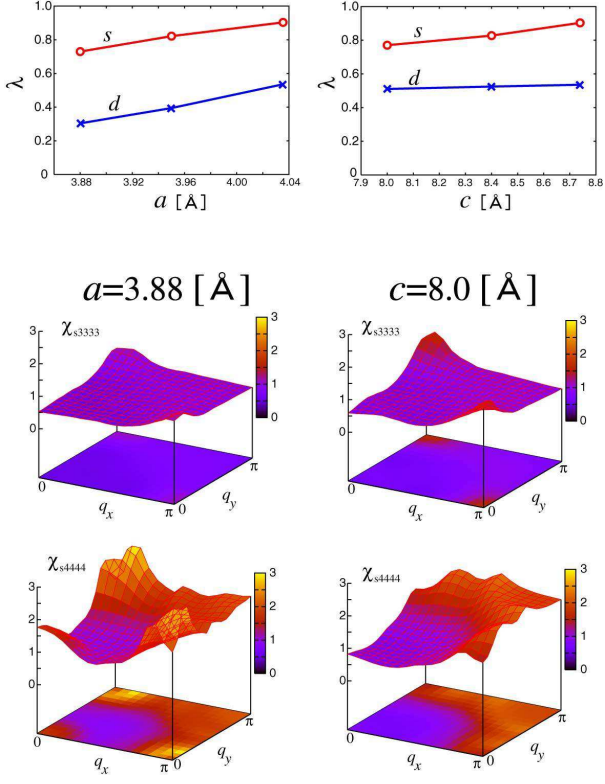}
\caption{(Color online) 
Upper panels : $s$-wave and $d$-wave 
eigenvalues of the Eliashberg equation 
plotted against $a$ (left) or $c$ (right), where $h_{\rm As}$ is 
fixed at the original value for LaFeAsO.  
Lower panels: $\chi_{s3333}$ and 
$\chi_{s4444}$ for $a=3.88$\AA \ with $c$ fixed at the original value (left) 
and $c=8.0$\AA \ with $a$ fixed at the original value (right).
\label{fig11}}
\end{center}
\end{figure}

The eigenvalue of the Eliashberg equation 
is plotted as functions of the lattice constants in
Fig.\ref{fig11}. We find that the 
reduction in the lattice constants tends to suppress superconductivity, 
which can be attributed to the increased hopping integrals and 
associated suppression of the electron correlation. 
In fact, the reduction in $a$ ($c$) enhances the $XZ,YZ$ ($X^2-Y^2$) 
hopping integrals, which leads to suppressed $\chi_{s3333}$ 
$(\chi_{s4444})$ as seen from the comparison between Fig.\ref{fig3} and 
the lower panels of Fig.\ref{fig11}. 
The effect of reduced lattice constants is small 
for the competition between $s$ and $d$ waves 
(i.e., two curves move roughly in parallel). 
We note here that, although an increased $h_{\rm As}$ and 
a decreased lattice constant 
$a$ both lead to a reduction in the Fe-As-Fe  
bond angle $\alpha$, they have opposite effects on the 
eigenvalue of the Eliashberg equation (compare Figs.\ref{fig9} and 
\ref{fig11}). 

\subsection{Effect of the orbital-dependent interactions}
\label{nakamura}
In refs.\onlinecite{Nakamura,Miyake}, it is pointed out that 
the interaction parameters have significant orbital dependence. 
In ref.\onlinecite{Nakamura}, the intraorbital 
repulsions are 
$U=3.27$, $2.77$, $2.20$, and $3.31$ (eV) for 
$d_{3Z^2-R^2}$, $d_{XZ/YZ}$, $d_{X^2-Y^2}$,  and $d_{XY}$ 
orbitals, respectively. 
This variation comes from the fact that each Fe 3$d$
orbital hybridizes with As 4$p$ quite differently.
Namely, while the local basis of the five-band model, $\{ d_i^\dagger \}$,  
can be represented as a linear combination
of atomic ${\tilde d^\dagger}$ and ${\tilde p}^\dagger$ orbitals
(as $\alpha_i {\tilde d}^\dagger_i + \sum_j \beta_{ij} {\tilde p}^\dagger_j$), 
the coefficients $\alpha_i, \beta_{ij}$ have a strong orbital dependence.
For example, the ratio $\beta_{ij}/\alpha_i$ is 
large for $i={X^2-Y^2}$ but small for e.g. $i=3Z^2-R^2$.  
Therefore, if we adopt a common value
for the interaction parameters for all five orbitals in the five-band model, 
electron correlations are relatively overestimated for $X^2-Y^2$,
because ${\tilde p}$ orbitals are more weakly correlated than 
${\tilde d}$ orbitals. In order to avoid this problem, 
we should use orbital dependent 
interactions, where the interaction for $i=X^2-Y^2$ is small compared to 
others.

So we study in this section the effect of the orbital 
dependence of the interactions, taking into account 
the orbital dependence of 
$U'$, $J$, $J'$ as well. Since the self-energy correction is not taken 
into account in RPA, it is again necessary to reduce the 
interactions to avoid magnetic ordering at high 
temperatures.  Here we multiply all the interaction parameters in 
ref.\onlinecite{Nakamura} by a factor of $f=0.42$, 
so the intraorbital interaction is taken 
to be 1.37, 1.16, 0.92, and 1.39 (eV) for 
$d_{3Z^2-R^2}$, $d_{XZ/YZ}$, $d_{X^2-Y^2}$,  and $d_{XY}$ orbitals,
respectively.
Since the $d_{X^2-Y^2}$ orbital has the smallest 
intraorbital interaction (0.92eV in the present calculation as compared 
with 1.2eV in the calculation for orbital-independent interactions), 
the effect of the $d_{X^2-Y^2}$ orbital is expected to be 
reduced compared with the results obtained by using the 
orbital independent interactions.

\begin{figure}[h]
\begin{center}
\includegraphics[width=8.0cm,clip]{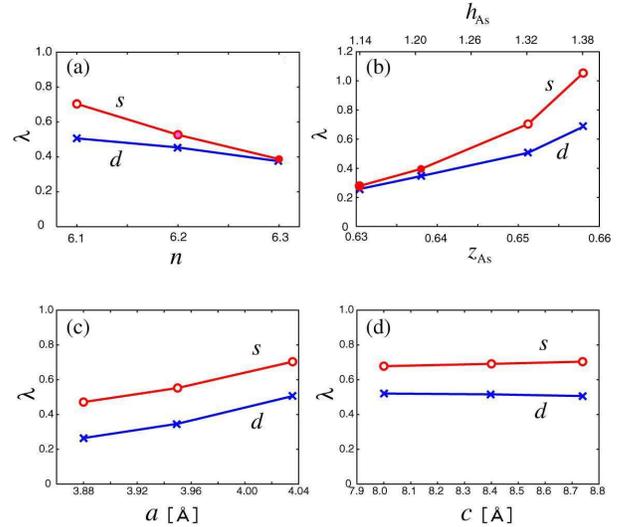}
\caption{(Color online) 
$s$-wave and $d$-wave eigenvalues of the Eliashberg equation
calculated for orbital-dependent interactions, 
plotted against (a) $n$ for the model of LaFeAsO, (b) $z_{\rm As}$ 
or $h_{\rm As}$ for $n=6.1$ with the lattice constants fixed at the original 
values for LaFeAsO, (c) $a$, and (d) $c$ with $h_{\rm As}$ fixed at the 
original value for LaFeAsO.  In (a), the light red (or gray) symbol 
for $s$-wave at $n=6.2$ indicates that the gap is nearly nodal, while 
the red (or solid) symbol at $n=6.3$ stands for the nodal $s$-wave.
\label{fig12}}
\end{center}
\end{figure}

First, we take the model for LaFeAsO to study 
the band-filling dependence. As shown in Fig.\ref{fig12}(a), 
we find that the $s$-wave pairing becomes 
nodal for $n>6.2$, i.e., the $s$-wave becomes nodal for smaller 
electron doping compared to the case with orbital 
independent interactions (see Fig.\ref{fig6}(a)). Also, the nodal 
$s$-wave pairing still slightly dominates over $d$-wave even at 
$n=6.3$, at which,  for the case of orbital-independent interactions, 
the $s$ gives way to $d$.
If we turn to the pnictogen-height dependence of the 
eigenvalue of the Eliashberg equation in Fig.\ref{fig12}(b), 
the $s$-wave is 
again enhanced with the increased height, but the enhancement is 
smaller than in the case of Fig.\ref{fig9}, which can be attributed to the 
reduction in the $\gamma$ (namely, the $d_{X^2-Y^2}$) Fermi surface effect).
As for the lattice constant dependence depicted in Fig.\ref{fig12}(c)(d), 
we find that the reduction in $a$ suppresses $\lambda$, while 
that of $c$ has small effect. This is because the reduction in  
$c$ mainly enhances the $d_{X^2-Y^2}$ hopping, 
and the suppression of the electron correlation within this 
orbital has small effect when the 
intraorbital interaction is small. 

\begin{figure}[h]
\begin{center}
\includegraphics[width=8.0cm,clip]{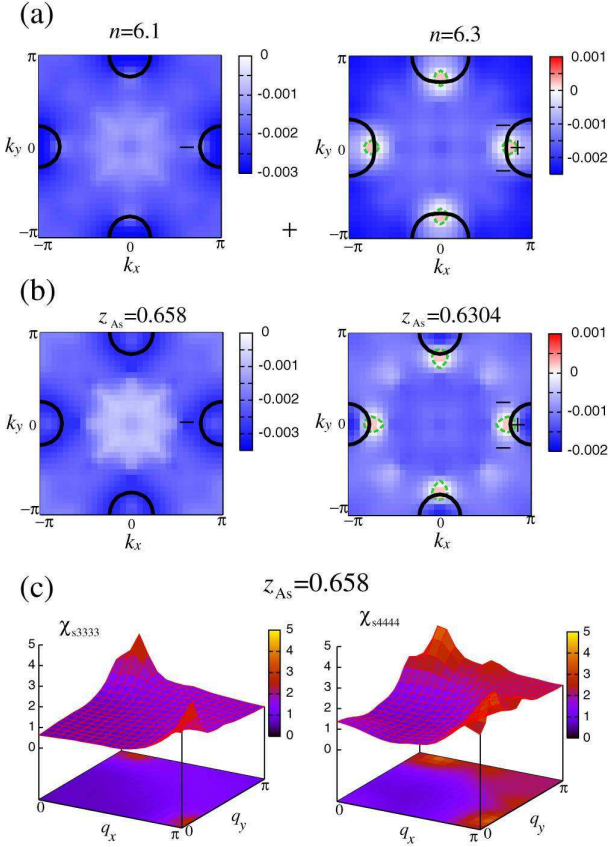}
\caption{(Color online) 
The gap function for band 4 
calculated with the orbital-dependent interactions 
for (a) LaFeAsO with $n=6.1$ (left) and $n=6.3$ (right), and  
(b) $z_{\rm As}=0.658$ (left) and $z_{\rm As}=0.6304$ (right) with 
$n=6.1$. (c) $\chi_{s3333}$ and $\chi_{s4444}$ for $z_{\rm As}=0.658$ and 
$n=6.1$.
\label{fig13}}
\end{center}
\end{figure}

Another effect of adopting orbital-dependent 
interactions appears in the symmetry of the gap function. 
For $z_{\rm As}=0.658$ with $n=6.1$, we have seen that 
the gap is large at the $d_{X^2-Y^2}$ charactered 
portions of the Fermi surface   
when we adopt orbital-independent interactions.  
For the orbital-dependent interactions, the absolute value of the 
gap is nearly constant for each of the all pockets
as shown in the left panel of Fig.\ref{fig13}(b) for band 4.
This should be again because the magnitude of the gap is reduced at 
$d_{X^2-Y^2}$ portions of the Fermi surface due to the reduction in the 
$d_{X^2-Y^2}$ intraorbital interaction. The effect of reducing 
the  $d_{X^2-Y^2}$
orbital interaction can be clearly seen in the comparison between 
$\chi_{s3333}$ and $\chi_{s4444}$ depicted in 
Fig.\ref{fig13}(c), where the two components have similar magnitudes for 
$z_{\rm As}=0.658$ in contrast to the result for the orbital-independent 
interactions in Fig.\ref{fig8}(a).

\section{Calculation for actual materials}
In this section, we calculate the band structure of 
actual materials other than LaFeAsO, i.e., the phosphate and 
Nd compound, 
using the experimentally determined lattice structure 
to construct the five-band model.  
The band filling will be fixed mainly 
at $n=6.1$ to make a direct comparison with the results for LaFeAsO.
 The results are interpreted in view of the
general trend obtained in the study of the virtual lattice structures.

\subsection{LaFePO}
\label{lafepo}
\begin{figure}[h]
\begin{center}
\includegraphics[width=8.0cm,clip]{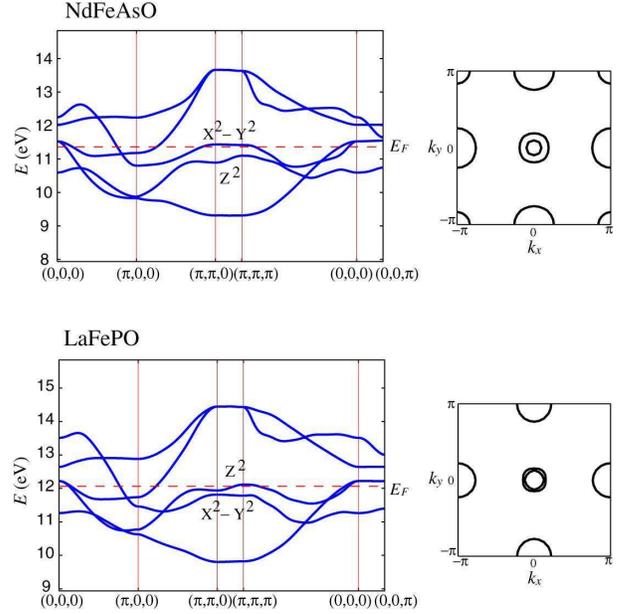}
\caption{(Color online) 
The band structure of the five-band model of the optimally doped 
NdFeAsO (upper panels) and 
LaFePO (lower). The Fermi surface at $k_z=0$ for $n=6.1$ is shown on the right.
\label{fig14}}
\end{center}
\end{figure}

\begin{figure}[h]
\begin{center}
\includegraphics[width=8.0cm,clip]{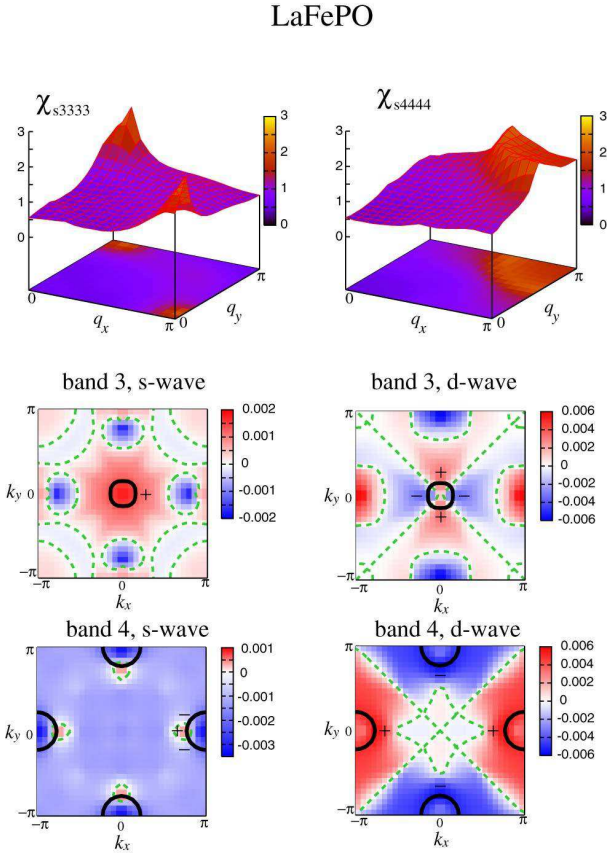}
\caption{(Color online) 
Upper panels : $\chi_{s3333}$ and $\chi_{s4444}$ 
for the five-band model for LaFePO with $n=6.1$, $U=1.7$, 
$U'=1.4$, and $J=J'=0.15$ (orbital-independent interactions). Lower panels : 
$s$-wave (left) and $d$-wave (right) gap functions for bands 3 and 4 
for the same parameter values.
\label{fig15}}
\end{center}
\end{figure}

The band structure of the five-band model for LaFePO is shown in
Fig.\ref{fig14}. 
In the case of LaFePO, the lattice constants are small compared to 
LaFeAsO (while closer to NdFeAsO below).  However, the hopping integrals are 
similar to or larger than those for the virtual structure for 
LaFeAsO with $z_{\rm As}=0.6304$. 
Thus, the main difference from LaFeAsO is caused by 
the height of P. The top of the $d_{X^2-Y^2}$ band at $(\pi,\pi)$ 
is indeed pushed below 
the Fermi level,\cite{GeorgesArita} 
and this makes the $(\pi,\pi/2)$ spin 
fluctuations arising from the $\beta$-$\beta$ nesting 
dominate in $\chi_{s4444}$ as shown in Fig.\ref{fig15}. 

\begin{figure}[h]
\begin{center}
\includegraphics[width=8.0cm,clip]{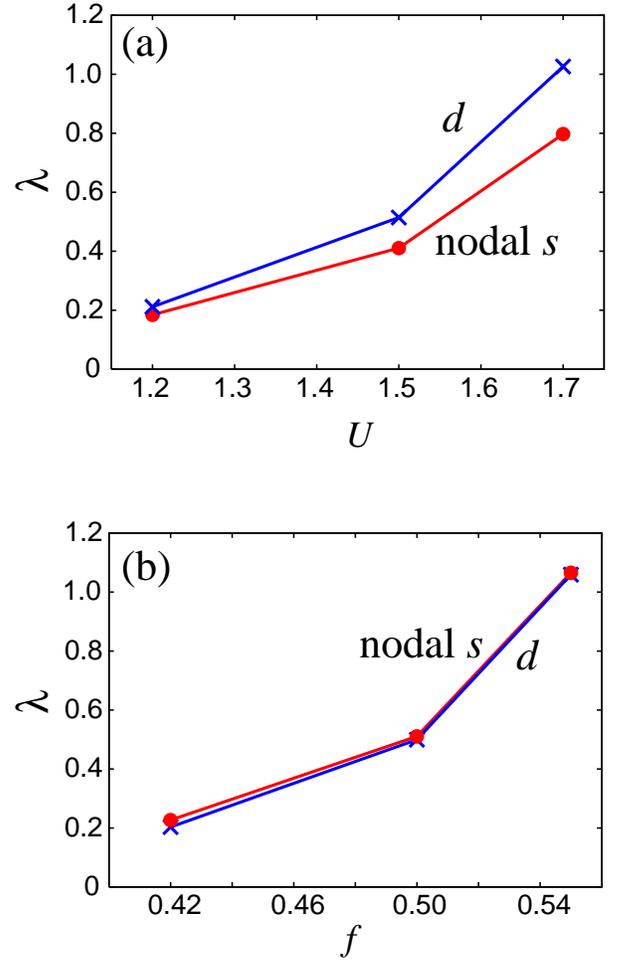}
\caption{(Color online) 
$s$-wave and $d$-wave eigenvalues of the Eliashberg equation 
for LaFePO plotted against (a) $U$ with orbital-independent interactions with 
$U-U'=2J=0.3$ fixed, and (b) 
the multiplication factor $f$ with orbital-dependent 
interactions. The $s$-wave gap here always has 
nodes intersecting the $\beta$ Fermi surface.
\label{fig16}}
\end{center}
\end{figure}

This behavior in the spin fluctuation 
for LaFePO 
acts to make the $d$-wave pairing (Fig.\ref{fig15} right) dominate 
for the orbital-independent interactions, while 
the sign-reversing $s$-wave with nodes 
intersecting the $\beta$ Fermi surface (Fig.\ref{fig15} left) 
is also closely competing.
The $U$ dependence of the eigenvalue of the 
Eliashberg equation is shown in Fig.\ref{fig16}(a). 
If we adopt the orbital-dependent interactions 
introduced in section\ref{nakamura}, on the other hand, 
we find that the nodal $s$-wave slightly 
dominates over $d$-wave in the 
entire parameter regime studied, as depicted 
in Fig.\ref{fig16}(b) for the eigenvalue of the Eliashberg 
equation against the interaction strength (i.e., the 
multiplication factor $f$ here). 
This is expected from the comparison between Fig.\ref{fig6}(a) 
and Fig.\ref{fig12}(a), 
where we can see that the orbital-dependent interaction tends to favor nodal 
$s$-wave over $d$-wave.

We further find that the $s$-$d$ competition depends on the 
band filling (not shown), i.e., 
smaller band fillings tend to favor the nodal $s$-wave.  
As seen from these results 
the competition between nodal $s$-wave and $d$-wave pairings in 
LaFePO is rather subtle, and it is difficult to theoretically determine which 
symmetry actually takes place.  In either case, however, 
superconducting gap of LaFePO is expected to have nodes intersecting the 
Fermi surface. This is in fact consistent with recent experiments on 
LaFePO that suggest the presence of nodes in the superconducting gap.
\cite{Fletcher,Hicks}

\subsection{NdFeAsO}
\label{ndfeaso}
\begin{figure}[h]
\begin{center}
\includegraphics[width=8.0cm,clip]{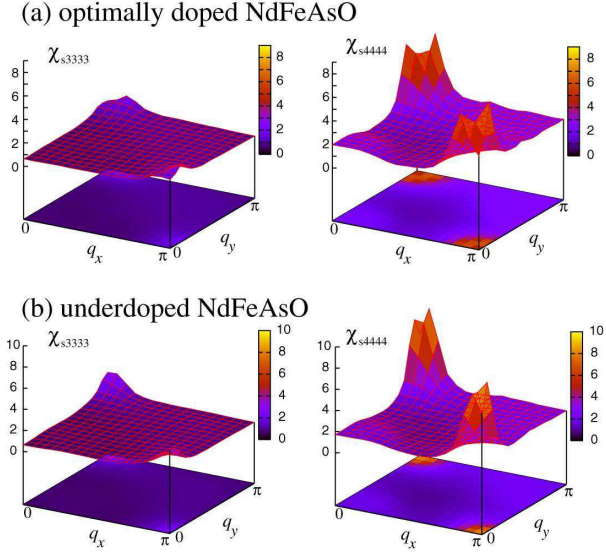}
\caption{(Color online) 
$\chi_{s3333}$ (left panels) and $\chi_{s4444}$ (right) in the five-band 
model for (a) the optimally doped NdFeAsO with $n=6.1$, and  
(b) the underdoped NdFeAsO with $n=6.03$.
\label{fig17}}
\end{center}
\end{figure}

\begin{figure}[h]
\begin{center}
\includegraphics[width=8.0cm,clip]{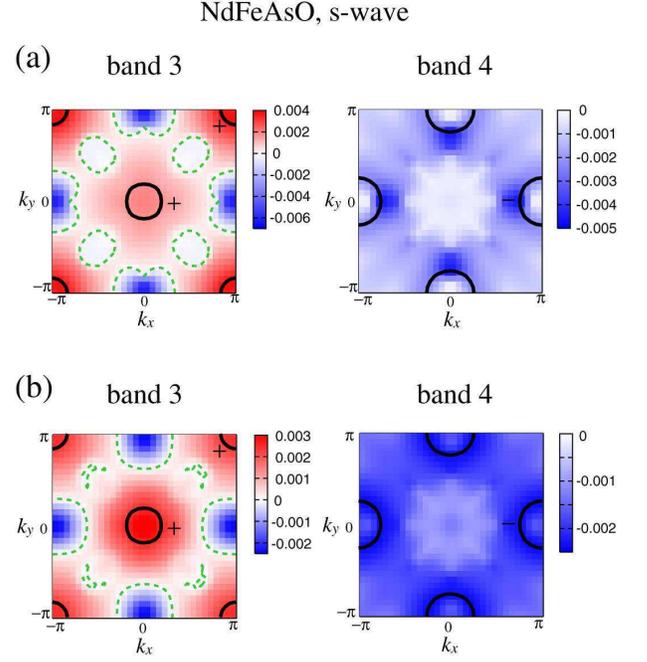}
\caption{(Color online) 
The $s$-wave gap function for band 3 (left panels) and band 4 (right) 
for the optimally doped 
NdFeAsO. Orbital-independent (a) and dependent (b) interactions are adopted.
\label{fig18}}
\end{center}
\end{figure}

The band structure of NdFeAsO is shown in Fig.\ref{fig14}. 
The low-temperature lattice structure 
of the optimally doped sample (sample 4) in ref.\onlinecite{Lee}
is adopted here.
Here, we have performed an LDA calculation with the plane wave basis set 
using the pseudopotential of Nd obtained with the 
open-core treatment for the $f$ electrons.
The $d_{X^2-Y^2}$ band at $(\pi,\pi)$ is seen to cross the 
Fermi level even at $n=6.1$ as expected, 
and the $(\pi,0)$ spin fluctuation strongly dominates in 
$\chi_{s4444}$ as shown in the right panel of Fig.\ref{fig17}(a), 
and the fully-gapped, sign-reversing $s$-wave 
shown in Fig.\ref{fig18}(a) strongly dominates over $d$-wave. 
For the orbital-independent interactions, 
the eigenvalue of the Eliashberg equation 
is $\lambda=1.20$  for $s$-wave, which 
is indeed greater than that for LaFeAsO ($\lambda=0.90$), 
but not as large as the 
virtual lattice structure of LaFeAsO 
where the As height is increased to the value of NdFeAsO ($\lambda=1.91$).
The latter property can mainly be attributed to the reduction in the 
lattice constants $a$ and $c$ in NdFeAsO as compared to those of LaFeAsO.

When we adopt the orbital-dependent interactions introduced in 
section\ref{nakamura}, the enhancement of $\lambda$ from the LaFeAsO value 
is weaker than in the case of orbital-independent interactions, 
as expected from the previous discussion. Namely, 
$\lambda_{\rm La}=0.70$ and $\lambda_{\rm Nd}=0.72$ for the 
multiplication factor $f=0.42$ while 
$\lambda_{\rm La}=1.20$ and $\lambda_{\rm Nd}=1.32$ for $f=0.45$.
The magnitude of the gap for the $s$-wave pairing is nearly 
constant on each of the all pockets as shown in Fig.\ref{fig18}(b), 
which is also expected from the argument in section\ref{nakamura}.

We have also performed a calculation for NdFeAsO with the lattice 
structure in the underdoped regime (sample 1 
in ref.\onlinecite{Lee}). Here we take the band filling of 
$n=6.03$ and adopt orbital-independent interactions.  
As shown in Fig.\ref{fig17}, 
the maximum value of $\chi_{s4444}$ 
(and also $\chi_s$, not shown) 
is larger than in the 
case of the optimally doped lattice structure
because the nesting is better for the underdoped case. 
Nonetheless, the eigenvalue of the Eliashberg equation is 
found to be smaller, $\lambda=1.03$ for the $s$-wave as compared to 
$\lambda=1.2$ for the optimally doped sample.
This shows that removing the electrons 
to make the $\gamma$ Fermi surface more effective does not necessarily 
favor superconductivity. This may be because lowering the 
Fermi level results in the decrease in the $d_{X^2-Y^2}$ density of 
states on the $\beta$ Fermi surface, where the $d_{X^2-Y^2}$ band 
forms a Dirac cone. Doping the electrons raises the Fermi level, thereby 
increasing the $d_{X^2-Y^2}$ density of states on the $\beta$ Fermi surface, 
and at the same time increasing the pnictogen height, which 
pushes up the $d_{X^2-Y^2}$ band 
at $(\pi,\pi)$  so as to catch up with the raised Fermi level. 
The increased $d_{X^2-Y^2}$ density of states upon 
doping is seen in the broad $(\pi,0)$ peak structure 
in $\chi_{s4444}$ in the optimally doped case as compared to that in the 
underdoped regime (Fig.\ref{fig17}).

The tendency that electron doping tends to increase the 
pnictogen height should be general because the increase in the negative charge 
in the Fe layers suppresses the attractive interaction between the 
positively charged iron and the negatively charged pnictogen. Therefore, the 
doping dependence of $\lambda$ shown 
in Fig.\ref{fig6}(a) with a fixed band structure  
may be too naive in that the effect of the $\gamma$ Fermi surface 
monotonically decreases with the higher electron doping.
As for the doping dependence of superconductivity, 
other than the effect of the change in the band structure,
there have also been theoretical studies that suggest the 
importance of the ``unscreening effect'' of the Coulomb interaction
\cite{Fuseya}, or the importance of the electron
correlation.\cite{Ikeda}

We have finally examined the effect of pressure on NdFeAsO. 
The lattice structure data for NdFeAsO$_{1-y}$ under a pressure of 
3.8GPa is taken from ref.\onlinecite{Kumai}.
Applying pressure on NdFeAsO tends to reduce $h_{\rm As}$ as well as the 
lattice constants. At 3.8GPa, the height is reduced to $h_{\rm As}=1.33$\AA. 
This results in a suppression of 
the eigenvalue of the Eliashberg equation, and the $s$-wave eigenvalue 
is reduced regardless of the choice of the electron-electron interactions, 
e.g., for the orbital-independent interactions, $\lambda$ at $T=0.02$ 
reduces (from 1.2) to 0.67.  The suppression of $\lambda$ 
is at least qualitatively consistent with the 
experiment.\cite{Takeshita,comment}

\section{Discussions}
\subsection{Validity of the five-band model}
One of the most important next steps in the microscopic study
on superconductivity in the iron-based superconductors 
should be an examination of the
present scenario based on the Fermi surface nesting by means of 
self-consistent calculations. Indeed, if we are interested in the 
behavior of the present five-orbital ($d$-only) model with moderate 
(realistic) size of the Hund coupling\cite{Anisimov} 
or the Coulomb interactions\cite{Nakamura,Miyake} 
estimated by various {\it ab initio} methods, 
we have obviously to take account of the self-energy 
(otherwise we have magnetic ordering at rather high temperatures), 
where the self-energy correction generally affects the effect of
Fermi surface nesting, which is usually overestimated in RPA.

One possible way to go beyond RPA is employing the
fluctuation exchange (FLEX) approximation\cite{Bickers}.
However, it has recently been recognized that FLEX does not work
so successfully for the five-orbital model with moderate 
correlations. Namely, while we naively expect that the model 
should have a strong instability for the stripe-type antiferromagnetic 
ordering for undoped LaFeAsO, the spin susceptibility in 
FLEX has a peak at $(\pi,\pi)$, which corresponds to the 
checkerboard-type antiferromagnetic instability.   
Even in the weakly correlated regime, 
Ikeda\cite{Ikeda} had to introduce artificial level shifts 
for $d_{z^2}$ and $d_{x^2-y^2}$ to the original 
LDA band in order to avoid a large $d_{z^2}$ Fermi surface.

These problems seem to 
come from the fact that the self-energy
correction generally has a strong orbital dependence in the
five-orbital model 
rather than from some problems in FLEX. 
Even in the simple Hartree approximation
for the paramagnetic case,
the band structure and the Fermi surface 
dramatically change from those in LDA 
due to the Hartree field ($\sim U\langle n_i \rangle$), 
since the filling of each of the five orbitals varies so differently 
(e.g., $n\sim 0.8(0.5)$ for $d_{z^2}(d_{x^2-y^2})$).

On the other hand, in the $dpp$ model that takes account 
of Fe 3$d$ and
As 4$p$ and O 2$p$\cite{GeorgesArita} for which five Fe 3$d$
orbitals are similarly filled, we do not have
such problems as far as we introduce a gap
(the so-called double-counting term $\Delta$)
that depends 
on the difference between the correlations in 
Fe 3$d$ and As 4$p$\cite{Mizokawa}.
In fact, this double-counting term in the $dpp$ model
makes the situation in the five-orbital model subtle.
For the $dpp$ model, we can safely assume that $\Delta$
does not have a serious orbital dependence.
On the other hand, if we translate $\Delta$
in terms of the five-orbital model, we have to 
assume that $\Delta$ has a non-trivial
orbital dependence, since each Fe 3$d$ hybridizes with
As 4$p$ differently (e.g., while the hybridization
between $d_{x^2-y^2}$ and As 4$p$ is strong, those 
between $d_{z^2}$ and As 4$p$ are weak\cite{GeorgesArita}).
This should be one reason why Ikeda had to introduce an 
orbital-dependent level shift in his FLEX calculation\cite{Ikeda}.

Therefore, we believe that it is impossible to obtain any 
meaningful results in self-consistent calculation for the five-orbital model
without considering the orbital-dependent double-counting term,
while we have still no guarantee that the double-counting term 
can really make the five-orbital model mimic the original $dpp$ model. 
An interesting observation is that 
the situation is in sharp contrast with the case of 
high $T_c$ cuprates. For cuprates, aside from 
the issue of the validity of 
the single-band Hubbard model or the $t$-$J$ model, 
we can naively expect that the 
self-consistent solutions of these models can at least 
describe Mott insulator, metallic state with strong antiferromagnetic
fluctuations, etc. On the other hand, for iron pnictides 
we have to seriously examine whether 
the five-orbital model indeed has a self-consistent solution with
the stripe-type antiferromagnetic instability.
This is why we consider this an important future problem.

\subsection{Phase diagram}

\begin{figure}[h]
\begin{center}
\includegraphics[width=8.0cm,clip]{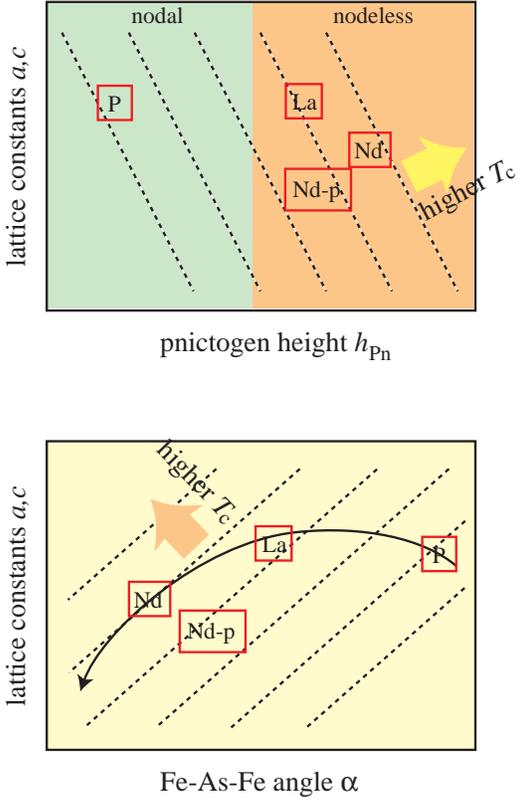}
\caption{(Color online) 
Schematic superconducting phase diagram on the $h_{\rm Pn}$-$[a,c]$ 
(upper) and  $\alpha$-$[a,c]$ (lower) planes. 
The dashed lines are schematic contours of $T_c$.
The curved arrow in the 
lower panel schematically indicates 
how the lattice parameters vary as the bond angle 
$\alpha$ is decreased in ref.\onlinecite{Lee}
The shorthands for materials are the same as in Table \ref{table1}.
\label{fig19}}
\end{center}
\end{figure}

In the preceding sections, it has been shown that the pnictogen 
height can act as a switch between high $T_c$ nodeless and low $T_c$ 
nodal superconductivities.  We have also shown that the increase in the 
lattice constants is unfavorable for superconductivity. 
These tendencies can be incorporated in a schematic 
phase diagram shown in Fig.\ref{fig19}. In the upper panel, we take the 
pnictogen height as the horizontal axis, and lattice constants 
as the vertical axis.  
In LnFeAsO, the lattice constants decrease monotonically 
in the chemical trend 
La$\rightarrow$ Nd$\rightarrow$ Dy.\cite{Miyazawa} On the other hand, the 
As height monotonically increases, and these effects may 
cancel with each other to result in a nearly constant $T_c$ between 
Ln=Nd and Dy.\cite{Miyazawa}
In the lower panel of Fig.\ref{fig19},
we adopt the As-Fe-As bond angle $\alpha$ as the horizontal 
axis to make clear comparison with  ref.\onlinecite{Lee}, in which 
it was shown that the maximum $T_c$ seems to be reached when the 
pnictogens form a regular tetrahedron.
As schematically indicated by a 
curved arrow, the appearance of the maximum $T_c$ may be a consequence of the 
combined effect of the bond angle and the lattice constants.
Thus, as far as the present 
theoretical study is concerned, the pnictogen height is a 
better parameter than the angle $\alpha$ to draw a phase diagram, 
since $\alpha$ is affected by both 
the height and the lattice constant $a$, which have opposite 
effects on $T_c$.

To attain higher $T_c$ on the basis of this phase diagram, 
it is desirable to have higher position of 
the pnictogen while keeping lattice constants not reduced.  
On the other hand, we have to keep in mind that 
such a variation in the lattice parameters also enhances the 
tendency toward magnetism.
Experimentally, magnetic ordering is known to occur 
after the structural phase 
transition, and the tendency toward magnetism and the structural 
phase transition seem to be linked.
In this sense, the superconductivity will have to compete with 
the magnetism/structural phase transition in a more 
severe manner as the height and/or the lattice constants are increased. 
Namely, too much increase in 
the height and/or the lattice constants can 
be unfavorable for superconductivity in that the 
magnetism/structural phase transition, which may take place 
at high temperatures, 
can dominate over superconductivity. 
In such a case, applying pressure to reduce the lattice constants 
and/or doping carriers can be effective to suppress the 
magnetism or the structural phase transition. 
This seems to be the case for the undoped
LaFeAsO\cite{Takahashi2} and CaFe$_2$As$_2$,\cite{Torikachvili} 
where structural 
phase transition and magnetism take place at ambient pressure, 
while applying pressure seems to remove the phase transition to 
result in superconductivity. 
A better understanding of the competition 
between superconductivity and magnetism should require 
further understanding of the magnetic state sitting next to the 
superconducting state. Namely, the mutual relation among the 
so-called ``spin density wave'' state observed experimentally
\cite{Dong,Cruz}, the magnetic state obtained in first 
principles calculations\cite{Dong,Mazin2,Ishibashi,Yildrim}, and 
the spin fluctuation obtained in the downfolded five-band models 
has to be made clearer.

Another point that should be kept in mind 
is the material dependence of the 
electron-electron interactions, which is not considered in the 
present phase diagram. In particular, the difference in the lattice 
structure in the 11 systems such as FeSe\cite{FeSe} and in the 
hole doped 
122 systems such as BaFe$_2$As$_2$\cite{Rotter}  can affect the effective 
electron-electron interaction within the FeAs planes. 
Also, the screening effect of the $f$ orbitals may be different 
between LaFeAsO and NdFeAsO. On the $f$ electrons, 
the present analysis adopts open-core treatment for the 
$f$ electrons of Nd, whose validity may have to be examined.  
We cannot completely rule out 
the possibility that the hybridization between Nd$f$ and Fe$d$ electrons 
can affect the electronic states and thus the superconductivity. 
In these senses, the actual phase diagram for the entire family of the 
iron-based superconductors should have more axes than presented 
here.

\section{Conclusion}

In the present study, we have investigated how the 
lattice structure affects the spin-fluctuation mediated
superconductivity in iron pnictides. 
The obtained picture is that the gap function and $T_c$ 
are determined by the competition or cooperation of the 
multiple spin fluctuation modes arising from $\alpha$-$\beta$, 
$\beta_1$-$\beta_2$, and $\gamma$-$\beta$ Fermi surface nestings, 
which depends on the materials, band filling, and/or pressure.
In particular, the competition between $\beta_1$-$\beta_2$ and 
$\gamma$-$\beta$ nestings within portions of the Fermi surface 
having strong $d_{X^2-Y^2}$ character governs the form of the 
superconducting gap as well as the strength of the superconducting 
instability. The relevance of the $\gamma$ (quasi) Fermi surface 
is determined by the pnictogen height, and consequently, the 
pnictogen height plays the role of a switch between 
high-$T_c$, nodeless and low-$T_c$, nodal pairings, which may give the 
answer to the question of why the form of the superconducting gap as well as 
$T_c$ are vastly different between LaFeAsO and LaFePO.\cite{Fletcher,Hicks}
An intriguing observation is, since $d$-wave and the nodal $s$-wave 
tend to be closely competing for 
low pnictogen heights, there is a possibility of 
exotic pairing such as $s+id$.\cite{Zhang}
The lattice constant also affects superconductivity in the 
manner that the reduction in $a$ ($c$) mainly suppresses the 
electron correlation within $d_{XZ}/d_{YZ}$ ($d_{X^2-Y^2}$) orbitals and 
thus degrades superconductivity. A schematic phase diagram has 
been obtained 
by combining the effect of the pnictogen height and the lattice constants. 

The low $T_c$ in cases where the $\beta_1$-$\beta_2$ nesting dominates 
over $\gamma$-$\beta$ nesting can be naturally understood in the sense that 
there is a kind of frustration 
between 
$\beta_1$-$\beta_2$, $\alpha$-$\beta_1$, and $\alpha$-$\beta_2$ nestings 
in determining the form of the gap because the sign of the gap 
has to be changed across each of the multiple nesting vectors 
(see Fig.\ref{fig4}). 
Conversely, the high $T_c$ in the case where the 
$\gamma$-$\beta$ nesting, along with $\alpha$-$\beta$, 
dominates over $\beta_1$-$\beta_2$ is 
natural in that the unfrustrated 
gap fully opens on all five disconnected 
pieces of the Fermi surface, 
$\alpha_1$, $\alpha_2$, $\beta_1$, $\beta_2$, and $\gamma$. 
In this sense, high $T_c$ 
in iron pnictides can be understood as a 
realization of the theoretical proposal 
that one can look for high $T_c$ superconductors in 
systems with disconnected Fermi surfaces.
\cite{KurokiDisconn,KurokiDisconn2}

The form of the $s$-wave gap is shown to be nonuniversal, even 
when the gap is fully open, and its variation 
along the Fermi surface strongly depends on the band structure 
(i.e. the lattice structure), 
the band filling, and the electron-electron  
interactions. When the $s$-wave gap 
varies significantly along the Fermi surface, it is also expected to be 
affected by the presence of the impurities, as pointed out by 
Mishra {\it et al}.\cite{Mishra}
In this sense, the form of the gap should experimentally be  
determined by a combination of multiple experiments on the same material,
desirably on the same sample. From this viewpoint, the 
discrepancy between the NMR experiments for LaFeAsO and other 
experiments suggesting nearly isotropic gap may be a consequence of the 
nonuniversality of the superconducting gap, especially because 
LaFeAsO lies close to the nodeless/nodal boundary.
In fact, anisotropic $s\pm$wave pairing where the gap 
varies strongly on the $\beta$ Fermi surface has been proposed 
to explain the $T^3$ decay in the NMR experiment.\cite{Nagai}
This view may also give some clue as to why some of the tunneling 
spectroscopy measurements exhibit zero bias conductivity peak,
\cite{Shantunnel} 
which is an indication of 
unconventional sign reversing pairing,\cite{Kashiwaya}
while others do not.\cite{Chentunnel} 
It is worth noting that recent theoretical 
studies show that it is unlikely to observe the 
zero bias conductivity peak for the fully gapped sign reversing 
$s$-wave pairing.\cite{Golubov,Onaritunnel}
 
In the present study, we have focused on the 
LnFeAsO (1111) systems.
In the 11 systems such as FeSe, experiments under pressure also suggest strong 
structure dependence of superconductivity.
\cite{Mizuguchi,Margadonna2,Cava,Masaki}
However, there may be some discrepancies from the 1111 systems regarding the 
lattice structure dependence since the LnO layer is not present.
Also, the extremely high position of the chalcogen atom in 11 
systems (1.47\AA \ in FeSe\cite{Margadonna} 
and 1.76\AA \ in FeTe\cite{Mizuguchi2}) 
may affect the lattice structure dependence.
Our study focusing on the 11 system is now underway.
Similarly, the hole-doped 122 systems are also expected to have some 
discrepancies with the electron-doped 1111 systems, which also deserves
future study.

\acknowledgments
We wish to thank H. Eisaki, T. Ito, C.-H. Lee, R. Kumai, and N. Takeshita for 
valuable discussions on the correlation between the lattice structure and 
$T_c$, and also providing some of the lattice structure data prior to 
publication.  We also thank Y. Tanaka and H. Kontani for valuable discussions.
RA would like to thank 
K. Nakamura, M. Imada, A. Toschi, P. Hansmann, 
G. Sangiovanni, K. Held, Z. Pchelkina, I. Solovyev, and H. Ikeda  
for fruitful
discussion on the validity of the five-band model.
Numerical calculations were performed at the facilities of
the Information Technology Center, University of Tokyo, 
and also at the Supercomputer Center,
ISSP, University of Tokyo. 
This study has been supported by 
Grants-in-Aid for Scientific Research from  MEXT of Japan and from 
the Japan Society for the Promotion of Science. H.U. acknowledges support by 
the Japan Society for the Promotion of Science.

%



\begin{thebibliography}{99}
\bibitem{commentbeta} Although the $\beta_1$ and $\beta_2$ 
Fermi surfaces are nearly circles centered around $(\pi,0)$ and $(0,\pi)$, 
the nesting vector is not $(\pi,\pi)$, since a nesting vector 
relevant to spin fluctuations has to connect electrons and holes so 
that they are as depicted in Fig.\ref{fig4}.
\bibitem{comment} The suppression of $\lambda$ from 1.20 (ambient pressure) 
to 0.67 (3.8GPa) 
seems to be too large considering that $T_c=45$K at 3.8GPa experimentally.
\cite{Takeshita} 
This may be because the lattice structure under pressure is determined 
at room temperature\cite{Kumai}, while we adopt the low temperature 
lattice structure 
for ambient pressure\cite{Lee}. In fact, it is found in 
ref.\onlinecite{Lee} that 
$h_{\rm As}$ increases upon lowering the temperature for NdFeAsO$_{1-y}$.
\end{thebibliography}
\end{document}